\documentclass[letterpaper,journal]{IEEEtran}
\usepackage{amsmath,amsfonts}

\usepackage[linesnumbered,ruled]{algorithm2e}

\usepackage{array}
\usepackage[caption=false,font=small,labelfont=rm,textfont=rm]{subfig}
\usepackage{textcomp}
\usepackage{stfloats}
\usepackage{url}
\usepackage{verbatim}
\usepackage{graphicx}
\usepackage{cite}
\usepackage{multirow}
\usepackage{booktabs}
\usepackage{makecell}
\usepackage{amssymb} 
\usepackage{cases} 
\usepackage{hyperref}  
\usepackage{tabularx}
\hypersetup{colorlinks=true,
	linkcolor=blue,
	anchorcolor=blue,
	citecolor=blue}
\usepackage{booktabs}
\usepackage[table]{xcolor}

\usepackage{float}
\usepackage{threeparttable}

\begin{document}
	
	\title{Flexible Multi-Target Angular Emulation for Over-the-Air Testing of Large-Scale ISAC Base Stations: Principle and Experimental Verification }
	
	\author{Chunhui Li,  \textit{Graduate Student Member, IEEE}, Hao Sun, and  Wei Fan,\textit{ Senior Member, IEEE}
		\thanks{    This work was supported by the National Natural Science Foundation of
			China under Grant  62571119. \textit{(Corresponding author: Wei Fan.)}}

		\thanks{ Chunhui Li and Wei Fan are with the National Mobile Communications Research Laboratory,	School of Information Science and Engineering, Southeast University, Nanjing 210096, China, and also with the Purple Mountain Laboratories,  Nanjing 211111, China (e-mail: lichunhui@seu.edu.cn; weifan@seu.edu.cn). }

		\thanks{Hao Sun is with the National Mobile Communications Research Laboratory,	School of Information Science and Engineering, Southeast University, Nanjing 210096, China, and also with the China Academy of Information and Communications Technology, Beijing 100191, China. (e-mail: sunhao@caict.ac.cn) } 
		
	}
	
	
	
	\maketitle
	
	\begin{abstract}
		Over-the-air (OTA) emulation of diverse sensing target characteristics in a controlled laboratory environment is pivotal for advancing integrated sensing and communication (ISAC) technology, as it facilitates the non-invasive performance evaluation of ISAC base stations (BSs) across complex scenarios.
		In this work, a flexible  multi-target  OTA emulation framework based on a wireless cable method is proposed  to evaluate the sensing performance of large-scale ISAC BSs. 
		The core concept leverages an amplitude and phase modulation (APM) network to simultaneously establish wireless cables and simulate target spatial characteristics without consuming additional resources on costly radar target emulators.
		For the wireless cable method, the condition number increases as the number of antennas scales up, which affects the performance of the wireless cable. Although the wireless cable concept has been established for devices-under-test (DUTs) with a limited number of antenna ports, establishing wireless cables for large-scale DUTs remains an open question in the community. We address this problem by optimizing the OTA probe array configuration based on the theoretical properties of strictly diagonally dominant matrices.
		Experimental results validate the proposed framework, demonstrating high-isolation wireless cables for a 32-element DUT and an extremely low condition number for a 128-element synthetic array. Furthermore, the OTA emulation of a dynamic dual-drone scenario confirms the method's effectiveness and practicality in reproducing complex sensing environments.

	\end{abstract}
	

	\section{Introduction}
	\IEEEPARstart{I}{ntegrated} sensing and communication (ISAC) is recognized as a pivotal candidate technology for sixth-generation (6G) mobile communication systems wherein sensing and communication functionalities share a common hardware platform \cite{wang2023road,you2021towards}. By enabling the efficient sharing of multidimensional resources across spatial, temporal, and frequency domains, this technology empowers ISAC base stations (BSs) to simultaneously perform wireless communication and radar sensing to ensure the coexistence and mutual enhancement of both operations, thereby facilitating innovative application scenarios such as low-altitude drone surveillance, vehicular network assistance, digital twins, and localization support \cite{liu2022integrated}.
	
	\begin{figure}[!t]
		\centering
		\includegraphics[width=0.48\textwidth]{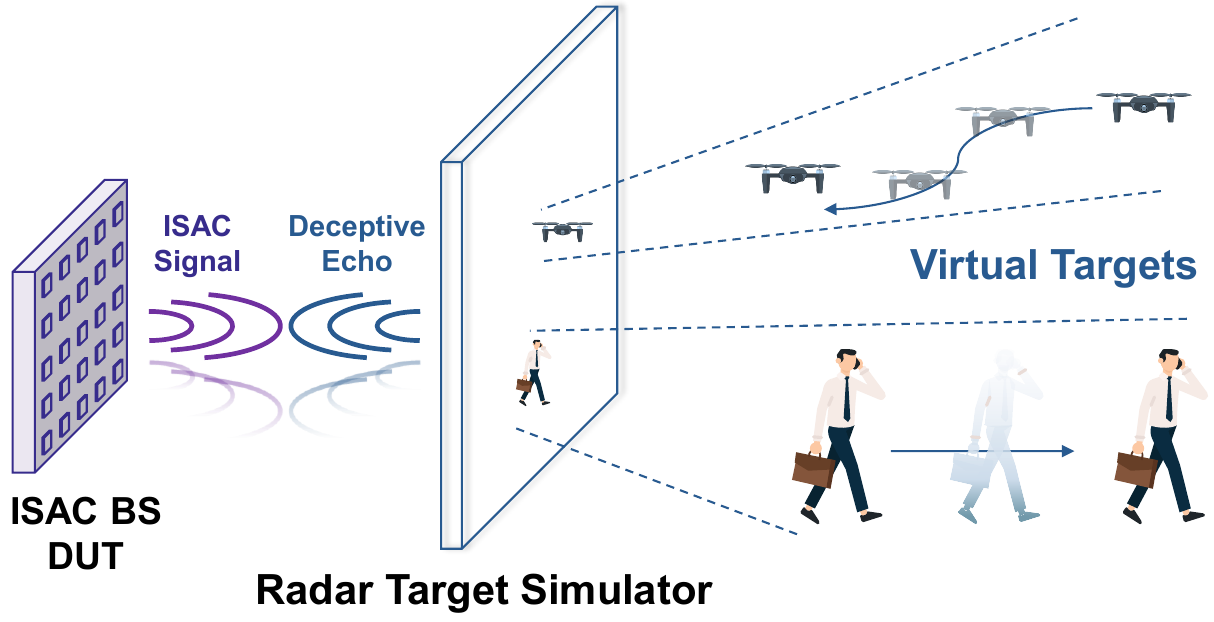} 
		\caption{Schematic diagram of the RTS simulating  targets for the  ISAC BS.
		}
		\label{fig_RTS_function} 
	\end{figure}
	
	To ensure that ISAC BSs can accurately estimate the angle of arrival (AoA), range, velocity, and radar cross section (RCS) of targets for real-world deployment, comprehensive sensing performance testing is indispensable \cite{liu2024senscap}. However, the introduction of ISAC BSs presents novel testing challenges as it necessitates the evaluation of sensing capabilities in addition to the conventional assessment of communication performance \cite{wang2025channel}. For the sensing verification of ISAC BSs, field testing serves as the commonly adopted approach utilizing real targets or equivalent physical models in realistic test environments to evaluate sensing capabilities \cite{zhang2022time,ding2025bi}. While this method is fundamental, it possesses inevitable limitations. First, field testing is time- and labor-intensive as ISAC BSs and sensing targets must be deployed and adjusted at designated sites often under non-laboratory conditions. 
	Second,  the high cost associated with constructing realistic test environments restricts the number of feasible test cases.
	Finally, field testing cannot guarantee a controllable electromagnetic environment, thereby limiting the repeatability of test cases.

	Given the aforementioned limitations of field testing, both academia and industry are actively pursuing more flexible testing approaches that enable the accurate evaluation of the sensing performance of ISAC BSs under realistic deployment conditions within controlled laboratory environments. 
	The sensing testing scheme based on radar target simulator (RTS) has attracted  research attention \cite{diewald2021radar}. As illustrated in Fig. \ref{fig_RTS_function}, this approach operates by employing an RTS to receive the ISAC signal from the device under test (DUT) ISAC BS and to generate and re-radiate deceptive echoes that emulate virtual targets characterized by specified AoA, ranges, velocities, and RCSs.
	Note that a commercial RTS module typically possesses the capability to simulate ranges, velocities, and RCSs, whereas the emulation of AoA necessitates the configuration of specific antenna array front ends implemented through distinct schemes.
	Within the radar community, many research has been investigated on RTS-based sensing test systems featuring diverse functionalities. Xing \textit{et al.} \cite{xingPhotonicsbasedBroadbandRadar2025} proposed a photonics-based broadband RTS designed to generate micro-Doppler signatures. For accurate moving target simulation, Körner \textit{et al.} \cite{korner2021multirate} proposed a multirate universal RTS. 
	Kern \textit{et al.} \cite{kernVirtuallyAugmentedRadar2024}  proposed an RTS designed for the training of radar machine learning algorithms.
	Additionally, the authors in \cite{wang2025channel} propose a cost-effective framework that repurposes a channel emulator (CE) to substitute for an RTS to achieve unified communication and sensing testing using a single instrument.
	
	Compared to the emulation of range and velocity characteristics for far-field point targets, the emulation of the AoA presents greater challenges and has emerged as a primary focus of recent research.
	Buddappagari \textit{et al.} \cite{buddappagari2021over} and Asghar \textit{et al.} \cite{asghar2021radar} implemented radar target simulators that utilize mechanical repositioning of radio frequency (RF) front-ends to synthesize the desired AoA. This methodology necessitates complex mechanical architectures and restricts the achievable target angular velocity. Alternatively, Scheiblhofer \textit{et al.} \cite{scheiblhofer2017low} and Gadringer \textit{et al.} \cite{gadringer2018radar} developed electronically switched architectures to avoid mechanical motion, yet these systems require a significant quantity of front-ends to maintain sufficient angular resolution. To mitigate these constraints and enable versatile emulation of spatial target characteristics, Diewald \textit{et al.} \cite{diewald2021arbitrary} introduced a RTS capable of arbitrary AoA synthesis through field synthesis techniques. 
	Moreover, Schroeder \textit{et al.} \cite{schoeder2021flexible,schoeder2023unified} proposed a flexible over-the-air (OTA) direction-of-arrival RTS by calibrating the propagation matrix that relates the RTS front ends to the antenna array of the DUT.
	In previous work \cite{liMultitargetFlexibleAngular2025}, a conducted multi-target flexible angular emulation RTS configuration was developed by integrating an amplitude and phase modulation network for sub-6 GHz ISAC BSs with accessible test ports.
	
	However, existing RTS schemes incorporating flexible AoA emulation are no longer suitable  for testing advanced ISAC BSs. On one hand, emerging Type-O ISAC BSs exhibit tightly integrated connections between the antenna array and RF chain module\cite{3GPP202xTS38141-1-tr}, resulting in the absence of accessible test ports, thereby rendering the conducted method in \cite{liMultitargetFlexibleAngular2025} inapplicable, especially for high frequency bands 
	On the other hand, although RTS schemes developed within the radar community operate in an OTA manner, they encounter challenges when adapted for ISAC BSs. These methods are primarily designed for automotive millimeter-wave (mmWave) radars operating in the V-band \cite{diewald2021arbitrary,schoeder2021flexible,schoeder2023unified}, typically adopting analog or hybrid beamforming architectures with limited RF chains.
	Conversely, ISAC BSs are intended to operate at lower frequencies, such as sub-6 GHz and emerging mid-bands, where digital or hybrid beamforming architectures with a large number of RF chains are expected to be utilized. It is also noteworthy that digital beamforming architectures for mmWave bands have seen significant progress in recent years.
	For performance testing, the critical requirement is the ability to guide desired testing signals to their respective receiver ports. However, existing methods fail to support DUTs equipped with a large number of RF chains and receivers. Specifically, for existing field synthesis-based RTS techniques \cite{diewald2021arbitrary}, achieving fine angular resolution necessitates densely deployed probe arrays. Similarly, for propagation matrix inversion-based RTS approaches \cite{schoeder2021flexible,schoeder2023unified}, the extensive number of BS antennas leads to a rapid increase in the condition number of the propagation matrix, which severely degrades emulation accuracy \cite{zhang2020achieving}. Consequently, OTA testing of BSs with large-scale antenna ports or receivers remains an open research question in the field. Therefore, this paper aims to address these gaps by developing a flexible multi-target angular OTA emulation method for large-scale advanced ISAC BSs lacking accessible test ports.

	\begin{table}[!t]
		\centering
		\begin{threeparttable}
			\caption{Comparison of Related RTS Approaches and This Work}
			\label{tab:radar-emulation-comparison}
			\scriptsize
			\setlength{\tabcolsep}{1.5pt}
			\renewcommand{\arraystretch}{1.3}
			
			\begin{tabularx}{\linewidth}{@{}p{1.15cm}p{2.1cm}ccXc@{}}
				\toprule
				\textbf{Ref.} & \textbf{Category} & \textbf{Ang. Flex.\tnote{1}} & \textbf{D\&D\tnote{2}} & \textbf{DUT Config.} & \textbf{Freq.} \\
				\midrule
				
				\cite{buddappagari2021over}, \cite{asghar2021radar} 
				& OTA (Mech. motion) 
				& Low 
				& Yes 
				& ARS 510 \newline (3Tx, 4Rx) 
				& 77, 79 GHz \\
				
				\cite{scheiblhofer2017low} 
				& OTA (Elec. switch) 
				& Low 
				& Yes 
				& ADF5904/1 \newline (1Tx, 4Rx) 
				& 24 GHz \\
				
				\cite{gadringer2018radar} 
				& OTA (Elec. switch) 
				& Low 
				& Yes 
				& ARS408 \newline (2Tx, 6Rx) 
				& 77 GHz \\
				
				\cite{diewald2021arbitrary} 
				& OTA (Field synth.) 
				& High 
				& Yes 
				& AWR1843 \newline (2Tx, 4Rx) 
				& 77 GHz \\
				
				\cite{schoeder2021flexible,schoeder2023unified} 
				& OTA (Tran. inv.) 
				& High 
				& Yes 
				& 1Tx, 4Rx \cite{schoeder2021flexible} \newline 6Rx \cite{schoeder2023unified} 
				& 78 GHz \\

				\cite{liMultitargetFlexibleAngular2025} 
				& Conducted 
				& High 
				& Yes 
				& 32 Tx, 32 Rx 
				& 3.5 GHz \\
				
				\textbf{This Work} 
				& \textbf{OTA  (Tran. inv.)  } 
				& \textbf{High} 
				& \textbf{Yes} 
				& \textbf{32 Tx, 32 Rx} 
				& \textbf{3.5 GHz} \\
				
				\bottomrule
			\end{tabularx}
			
			\begin{tablenotes}
				\item[1] Ang. Flex.: Angular Emulation Flexibility.
				\item[2] D\&D: Delay and Doppler Emulation.
				\item Abbreviations: Mech.: Mechanical; Elec.: Electronic; Tran. inv.: Transfer matrix inversion; Synth.: Synthesis; Config.: Configuration; Freq.: Frequency.
			\end{tablenotes}
		\end{threeparttable}
	\end{table}

	This work is inspired by the wireless cable method utilized in traditional communication testing \cite{yu2014radiated,fan2017mimo} and addresses the critical challenge of extending its applicability from small-scale to large-scale DUTs. The proposed method enables direct signal delivery from each probe antenna to the intended DUT antenna port without physical RF cable connections by measuring and calibrating the transfer matrix between the CE output ports and the DUT antenna ports, a concept that aligns with the approaches described in \cite{schoeder2021flexible,schoeder2023unified}.
	As previously discussed, this method is constrained by the condition number of the transfer matrix, posing significant challenges for DUT equipped with large-scale antenna arrays. 
	Related works have demonstrated implementations with four \cite{wangNovelWirelessCable2025,wangAchievingWirelessCable2025,schoeder2021flexible} to six \cite{schoeder2023unified} receive (Rx) antennas, configurations that fall short of meeting the OTA testing requirements for ISAC BS.
	In this work, a  wireless cable method is proposed by investigating the optimal array configuration and placement of the OTA probe array to minimize the condition number of the transfer  matrix for large-scale DUT. Integrated with an amplitude and phase modulation (APM) network and a RTS, this system enables flexible multi-target angular OTA emulation for the sensing assessment of ISAC BSs.
	A comparative summary of the proposed approach with existing relevant RTS schemes is provided in Table \ref{tab:radar-emulation-comparison}, and the main contributions of this work are outlined as follows.
	\begin{figure*}[!t]\centering
		\includegraphics[width=18 cm]{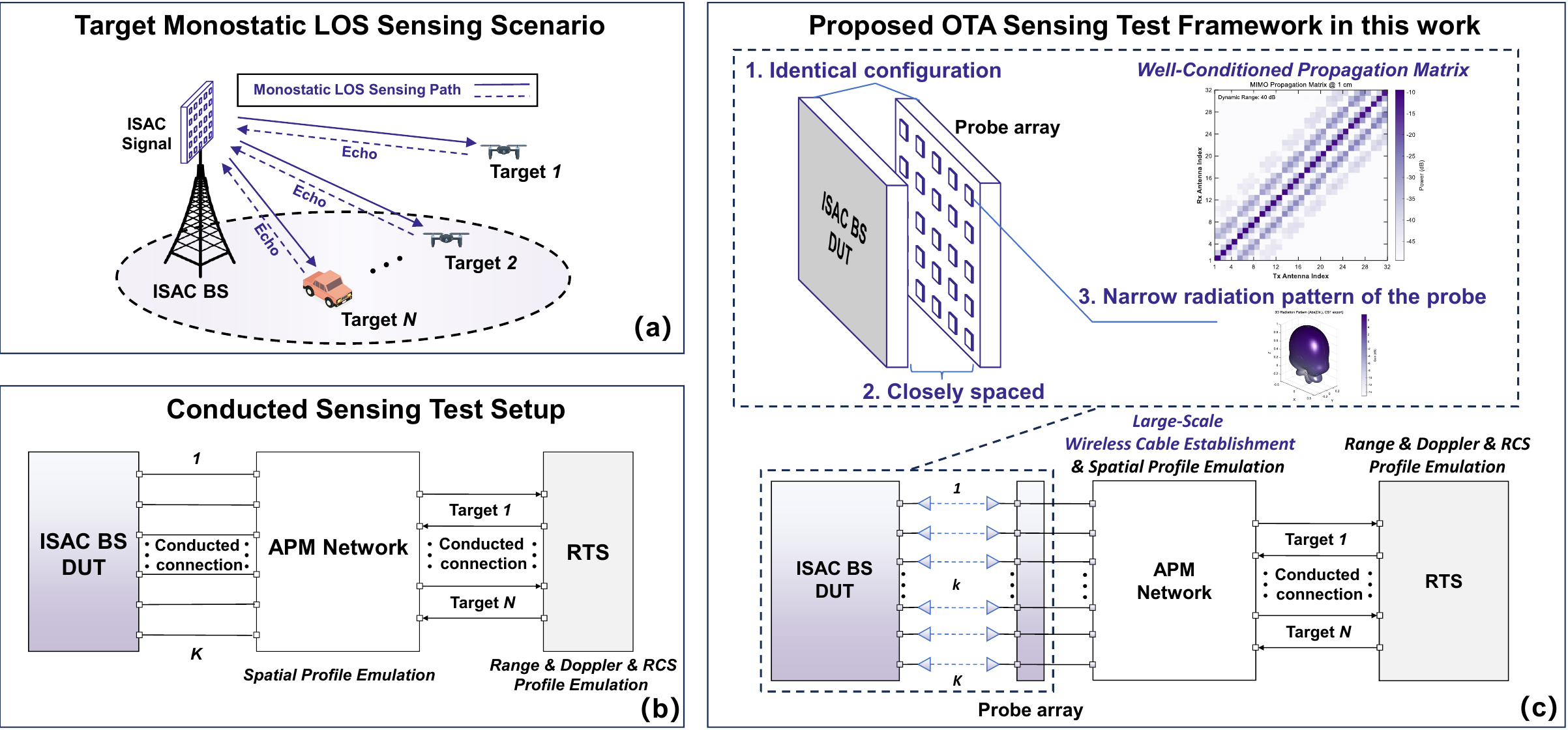}
		\caption{Overview of the  concept of this work. (a) A typical multi-target sensing scenario with a monostatic LOS sensing scheme. (b) The conventional conducted sensing test setup in \cite{liMultitargetFlexibleAngular2025}. (c) The proposed OTA sensing test framework.}
		\label{FIG_Framework}
	\end{figure*}
	
	\begin{itemize}
		\item 
		First, a flexible multi-target angular OTA emulation framework based on the large-scale wireless cable concept is proposed for ISAC BS sensing tests. The core idea is to deploy an APM network integrated with an OTA probe array between the DUT ISAC BS and an RTS. The APM network establishes the wireless cable and emulates the target angular profile, whereas the RTS emulates the target RCS, range, and Doppler profiles via its limited interface ports. The proposed framework enables OTA emulation of multiple targets with arbitrary RCS, range, angle, and Doppler profiles for ISAC BSs equipped with large-scale antenna arrays.
		This work is applicable to radar point target  in monostatic line-of-sight (LOS) sensing scenarios.
		
		\item 
		Additionally, establishing viable large-scale wireless cable links necessitates reducing the condition number of the propagation matrix between the OTA probe array and the DUT array. Therefore, based on the upper bound of the condition number for strictly diagonally dominant (SDD) matrices, we investigate the minimization of the propagation matrix condition number by optimizing the configuration and physical placement of the probe array.
		Several experiments are conducted to validate the proposed solution.  In the measured setup, the condition number of a $32 \times 32$ propagation matrix is reduced to $1.3$, and the average interlink isolation among wireless cable connections exceeds 30 dB. Under a virtually synthesized array, a $128 \times 128$ propagation matrix condition number as low as $3.2$ is achieved. To the best of our knowledge, this represents the highest array dimensionality reported for a practical wireless cable method in the literature.

		\item Finally, a representative drone sensing scenario is designed to validate the proposed framework, in which the sensing targets are experimentally emulated using both the proposed OTA framework and the conducted framework in \cite{liMultitargetFlexibleAngular2025}. The spatio-temporal-frequency characteristics of each drone are successfully estimated by the emulated ISAC BS under test in both the OTA  and conducted emulation architectures. The experimental results verify the feasibility of the proposed framework for OTA  emulation of multiple sensing targets in ISAC BS testing.
	\end{itemize}

	\textit{Notations}:
	Bold uppercase characters $\mathbf{X}$  denote matrices;
	bold lowercase characters $\mathbf{x}$ denote vectors;
	$(\cdot )^{{T}}$ is the transpose operator; $t$ and $f$ denote the time and frequency, respectively.

	\section{Problem Statement and Proposed Framework}
	This section states the problem and introduces the proposed sensing target OTA emulation framework. Furthermore, we formulate the signal models for both the target sensing scenario and the emulation system.

	\subsection{Problem Statement}
	Consider a typical monostatic ISAC scenario featuring LOS path sensing as illustrated in Fig. \ref{FIG_Framework} (a), wherein the ISAC BS is equipped with $K$ antenna elements to detect $N$ radar point targets.
	The RTS-based test setup aims to emulate the velocity, AoA, range, and RCS of these $N$ point targets for the DUT ISAC BS within a laboratory environment, thereby ensuring consistency in sensing conditions and performance between laboratory testing and real-world field deployments.
	In \cite{liMultitargetFlexibleAngular2025}, a conducted sensing test setup is proposed wherein the DUT ISAC BS, the APM network, and the RTS are sequentially interconnected via cables as shown in Fig. \ref{FIG_Framework} (b). The full-mesh APM network is responsible for emulating the target AoAs by simulating phase responses at each port of the DUT, while the RTS emulates the range, Doppler, and RCS characteristics. The advantage of this scheme is that the number of expensive RTS ports does not need to match the typically large number of antennas on the ISAC BS DUT but instead correlates only with the number of point radar targets.
	However, this conducted setup necessitates physical cabling between the ISAC BS and the APM network, which can require prohibitive time (approximately three
	days to one week in our case) for connection when dealing with massive multiple-input multiple-output (MIMO) BSs equipped with a large number of antennas. Massive cable connections are prone to errors, time-consuming, and susceptible to connector damage. Furthermore, emerging Type-O BSs lack accessible testing interfaces for physical cabling \cite{3GPP202xTS38141-1-tr}. 
	Consequently, the development of an OTA sensing target emulation scheme is the problem addressed in this work, which circumvents these issues by eliminating the need for physical connections.

	\subsection{Proposed Framework}
	In light of the limitations of  conducted schemes, our primary contributions are twofold. First, we propose an OTA emulation framework for ISAC BSs based on the wireless cable method, which has not been previously reported. Second, while the wireless cable concept has been introduced for MIMO terminal  and automotive antenna \cite{ji2019virtual}  performance testing, such as $2 \times 2$ \cite{li2023wideband} and $4 \times 4$ \cite{wangNovelWirelessCable2025,wangAchievingWirelessCable2025} MIMO with up to four receivers, and demonstrated for radar testing with few receivers, its applicability depends critically on the condition number of the transfer matrix. State-of-the-art works have indicated that the wireless cable method is unsuitable for DUTs with large-scale antenna ports \cite{fan2023wireless}. We address this critical limitation, rendering the wireless cable method viable for large-scale DUTs.
	We will introduce these two aspects sequentially.

	\subsubsection{Framework Composition} As illustrated in Fig. \ref{FIG_Framework} (c), a flexible OTA multi-target angular emulation framework based on the wireless cable concept is proposed. This framework comprises four components consisting of the DUT ISAC BS equipped with its integrated antenna array, the OTA probe array, the APM network, and the RTS.
	The OTA probe array serves to receive ISAC signals and reradiate deceptive echo signals toward the DUT, being connected to the APM network via cables. The APM network is employed to compensate for the transfer matrix between the DUT array and the OTA array to establish wireless cables, while also being responsible for simulating the spatial profile of each point target. Connected to the APM network, the RTS performs the same task as in the conducted scheme, which is to emulate the range, Doppler, and RCS characteristics of the point targets.
	
	In the wireless cable concept, the condition number of the transfer matrix between the DUT array and the OTA probe array antenna critically influences the isolation among the established wireless cable links. Within the proposed framework, the configuration and placement of the OTA probe array are of paramount importance as they constitute the primary degree of freedom (DoF) for optimizing the characteristics of the transfer matrix, which will be elaborated upon in subsequent sections.

	\subsubsection{Target Scenario Signal Model}
	Based on the monostatic sensing scenario illustrated in Fig. \ref{FIG_Framework} (a), we consider an ISAC BS equipped with a uniform planar array (UPA) of $K$ antennas operating in array duplex transmission and reception (ADTR) mode. The target channel frequency response (CFR) $\mathbf{H} (t,f) \in \mathbb{C}^{K \times K}$ comprises $N$ LoS channel components, which can be expressed as follows:
	\begin{equation} \label{eq_1}
		\mathbf{H} (t,f)=\sum_{n=1}^{N} \mathbf{H}_{n} (t,f).
	\end{equation}
	We adopt the point target assumption under far-field conditions in this paper. Thus, the $n$-th LoS channel component can be expressed as follows:
	\begin{align} \label{eq_2}
		\mathbf{H}_{n} (t,f) =  & G_{n}(t, \mathbf{\Theta}_{n}(t)) \cdot \operatorname{exp}(\mathrm{j} 2\pi \nu_{n}(t) t )\cdot \operatorname{exp}(\mathrm{j} 2\pi f \tau_{n}(t))   \nonumber
		\\
		&\cdot \mathbf{a}_{t}(\mathbf{\Theta}_{n}(t)) \cdot \mathbf{a}_{r}^{T}(\mathbf{\Theta}_{n}(t)), 
	\end{align}
	where $\mathbf{\Theta}_{n}(t)$ is the spatial direction vector of $n$-th sensing point target at time $t$.  
	$G_{n}(t, \mathbf{\Theta}_{n}(t))$ is the channel gain of the $n$-th channel component, determined by path loss, antenna pattern, and RCS. The term $G_{n}(t, \mathbf{\Theta}_{n}(t))$ can be obtained through various methods such as the  radar equation calculation or ray tracing.
	$\mathbf{a}_{r}(\mathbf{\Theta}_{n}(t)) \in \mathbb{C}^{K \times 1}$ and $\mathbf{a}_{t}(\mathbf{\Theta}_{n}(t)) \in \mathbb{C}^{K \times 1}$ are the receiver (Rx) and transmitter (Tx) array steering vectors in the direction of $\mathbf{\Theta}_{n}(t)$, respectively.
	Consider the ISAC BS operating in ADTR mode, where $\mathbf{a}_{r}(\mathbf{\Theta}_{n}(t))$ and $\mathbf{a}_{t}(\mathbf{\Theta}_{n}(t))$ are identical. Readers can refer to for the formulation of the steering vector of the UPA array.
	The Doppler shift $\nu_{n}(t)$ and delay $\tau_{n}(t)$ of the $n$-th channel component at time $t$ are determined by the target velocity and range, respectively.

	\begin{figure}[!t]
		\centering
		\includegraphics[width=0.49\textwidth]{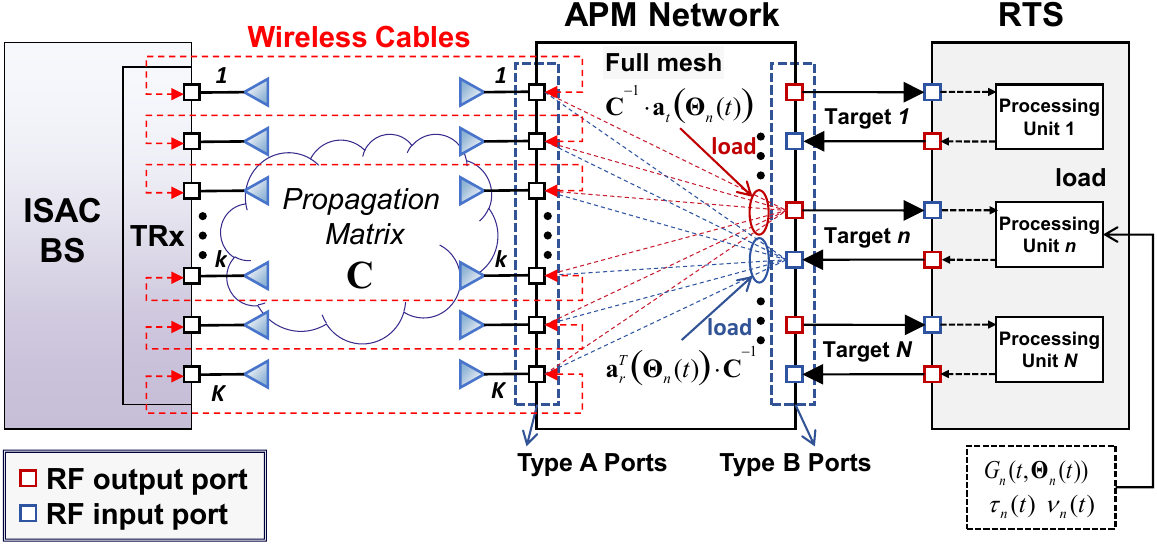} 
		\caption{Detailed signal model of the proposed framework.}
		\label{fig_ADTR} 
	\end{figure}

	\subsubsection{Framework Configuration and Signal Model}
	The objective of the proposed framework is to reconstruct the target field sensing channel (\ref{eq_2}) at the DUT ports.
	For the ISAC BS operating in ADTR mode, the specific configuration of the proposed framework is illustrated in Fig. \ref{fig_ADTR}. The RF ports of the APM network are segregated into two groups interconnected via internal links featuring adjustable amplitude and phase, designated as Type-A and Type-B ports, respectively. In this configuration, $K$ Type-A ports and $2N$ Type-B ports are activated to establish a full-mesh topology wherein the $k$-th Type-A port is connected to all $2N$ Type-B ports, and vice versa. The $K$ OTA probes are linked to the $K$ Type-A ports of the APM network via RF cables. 
	The Type-B ports are organized into $N$ groups with each group comprising one output port and one input port. This configuration is dictated by the architecture of the RTS employed for emulating $N$ targets.
	Within the $n$-th port group, the output Type-B port of the APM network transmits the signal to the RF input port of the $n$-th RTS processing unit via an RF cable.

	\begin{figure*} 
		\normalsize
		\begin{equation} \label{eq:long_equation1} 
			\hat{\mathbf{H}}_{n}=	\mathbf{C} \cdot 
			\underbrace{
				\hat{\mathbf{C}} ^{-1} 
				\cdot \mathbf{a}_{t}\bigl(\mathbf{\Theta}_{n}(t)\bigr)
			}_{\text{Emulated by the APM network}} \cdot 
			\underbrace{G_{n}\bigl(t, \mathbf{\Theta}_{n}(t)\bigr)
				\cdot
				\exp(\mathrm{j} 2\pi \nu_{n}(t) t )
				\cdot \exp(\mathrm{j} 2\pi f \tau_{n}(t))}_{\text{Emulated by the RTS}}
			\cdot 
			\underbrace{
				\mathbf{a}_{r}^{T}\bigl(\mathbf{\Theta}_{n}(t)\bigr)
				\cdot \hat{\mathbf{C}} ^{-1}}
			_{\text{Emulated by the APM network}}
			\cdot \mathbf{C} 
		\end{equation}
		\vspace*{4pt} 
		\begin{equation} \label{eq:long_equation2} 
			\hat{\mathbf{H}}=	\mathbf{C} \cdot \hat{\mathbf{C}} ^{-1} 
			\cdot \sum_{n=1}^{N} \left [ \mathbf{a}_{t}\bigl(\mathbf{\Theta}_{n}(t)\bigr) \cdot 
			G_{n}\bigl(t, \mathbf{\Theta}_{n}(t)\bigr)
			\cdot
			\exp(\mathrm{j} 2\pi \nu_{n}(t) t )
			\cdot \exp(\mathrm{j} 2\pi f \tau_{n}(t)) 
			\cdot \mathbf{a}_{r}^{T}\bigl(\mathbf{\Theta}_{n}(t)\bigr)
			\right ]  \cdot \hat{\mathbf{C}} ^{-1} \cdot \mathbf{C} 
		\end{equation}
		\vspace*{4pt} 
		\hrulefill
	\end{figure*}

	For the emulation of the $n$-th target, the emulated sensing channel $\hat{\mathbf{H}}_{n}$ based on the proposed configuration can be expressed as in (\ref{eq:long_equation1}), presented at the top of this page, which comprises the following components:
	\begin{enumerate}
		\item $\mathbf{C} \in \mathbb{C}^{K \times K}$ denotes the transfer matrix between the $K$ DUT antennas and the $K$ OTA probes. This transfer matrix encapsulates the radiation patterns of the DUT antennas and OTA probes, the propagation environment, and mutual coupling effects if the test environment is situated within near-field conditions. 
		\item  $\hat{\mathbf{C}} \in \mathbb{C}^{K \times K}$ represents the measured transfer matrix, which can be determined directly via an \footnotesize{ON--OFF} \normalsize procedure \footnote{The \scriptsize ON–OFF \footnotesize procedure entails a process wherein, during each measurement instance, only a single probe and one DUT antenna are on, while all other probes and DUT antennas remain off. Through this procedure, the transfer matrix between the probe ports and the DUT antenna ports can be acquired.} . Owing to real-world measurement inaccuracies, $\hat{\mathbf{C}}$ deviates from the true channel matrix $\mathbf{C}$ by an error matrix $\mathbf{E} \in \mathbb{C}^{K \times K}$, such that $\hat{\mathbf{C}} = \mathbf{C} + \mathbf{E}$. In general, the error matrix $\mathbf{E}$ is relatively small, satisfying $\left\| \mathbf{E} \right\|_{2} \ll \left\| \mathbf{C} \right\|_{2}$. $\hat{\mathbf{C}}^{-1} \in \mathbb{C}^{K \times K}$ serves as the calibration matrix for establishing wireless cables, and the condition $\mathbf{C} \cdot \hat{\mathbf{C}}^{-1} \approx \mathbf{I}$ indicates the successful establishment of the wireless cable link \footnote{This is because, in traditional conducted testing under ideal conditions, the antenna ports of the probe array and the DUT antenna array are directly interconnected via cables in a one-to-one configuration, corresponding to a transfer matrix that is an ideal identity matrix $\mathbf{I}$.}.

		\item $\mathbf{a}_{r}(\mathbf{\Theta}_{n}(t)) \in \mathbb{C}^{K \times 1}$ and $\mathbf{a}_{t}(\mathbf{\Theta}_{n}(t)) \in \mathbb{C}^{K \times 1}$ denote the array Rx and Tx steering vectors, respectively, as defined in (\ref{eq_2}). 
		The integrated component $\hat{\mathbf{C}}^{-1} \cdot \mathbf{a}_{t}\bigl(\mathbf{\Theta}_{n}(t)\bigr) \in \mathbb{C}^{K \times 1}$, employed to establish the wireless cable link and emulate the spatial characteristics of the Tx path, is loaded into the $K$ transmission links of the APM network for emulating the $n$-th target, as indicated by the red dashed lines in Fig. \ref{fig_ADTR}. 
		Similarly, the integrated component $\mathbf{a}_{r}^{T}\bigl(\mathbf{\Theta}_{n}(t)\bigr) \cdot \hat{\mathbf{C}}^{-1} \in \mathbb{C}^{1 \times K}$, employed to establish the wireless cable link and emulate the spatial characteristics of the Rx path, is loaded into the $K$ echo links of the APM network, as indicated by the blue dashed lines in Fig. \ref{fig_ADTR}.
		
		\item 
		The term $G_{n}\bigl(t, \mathbf{\Theta}_{n}(t)\bigr) \cdot \exp(\mathrm{j} 2\pi \nu_{n}(t) t ) \cdot \exp(\mathrm{j} 2\pi f \tau_{n}(t))$ encapsulates the RCS, range, and velocity information of the radar point target, which is emulated by the RTS. For a commercial RTS, the corresponding parameters can be directly configured for emulation. However, if a CE is employed for target emulation \footnote{The capability of a CE to function as a RTS has been validated in the literature \cite{wang2025channel,liMultitargetFlexibleAngular2025}.}, these radar parameters must be equivalently represented based on channel parameters such as power, delay and Doppler shift.
	\end{enumerate}

	For the proposed framework, the extension from single-target to multi-target emulation is straightforward. The emulated multi-target  sensing channel $\hat{\mathbf{H}}  \in \mathbb{C}^{K \times K}$ can be expressed as in (\ref{eq:long_equation2}), which is presented at the top of this page.

	\subsection{Discussion}
	The proposed framework is cost-effective, as the simulation links within the RTS are significantly more expensive than the APM links within the APM network. For the emulation of a single point target, only one internal RTS simulation link is required, alongside $2K$ APM links to match the number of DUT antennas.
	For multi-target simulation, the required resources increase linearly.
	
	To successfully apply the wireless cable concept within this framework, it is imperative to ensure that the condition $\mathbf{C} \cdot \hat{\mathbf{C}}^{-1} \approx \mathbf{I}$ is satisfied. Consequently, the propagation environment must remain time-invariant during the testing process to maintain a constant $\mathbf{C}$, as any variation necessitates the re-acquisition of $\hat{\mathbf{C}}$ to reconstruct the wireless cable link.
	Additionally, the most critical issue concerns the condition number of $\mathbf{C}$, as it determines the sensitivity of the matrix calculation to noise and errors, defined as
	\begin{equation} 
		\kappa(\mathbf{C})=\frac{\sigma_{\max }(\mathbf{C})}{\sigma_{\min }(\mathbf{C})}=\|\mathbf{C}\| \cdot\|\mathbf{C}^{-1}\|.
	\end{equation}
	
	A large condition number indicates that $\mathbf{C}$ is \textit{ill-conditioned}, meaning small perturbations in $\mathbf{C}$ can cause significant deviations in $\hat{\mathbf{C}}^{-1}$, thus preventing the ideal approximation $\mathbf{C} \cdot \hat{\mathbf{C}}^{-1} \approx \mathbf{I}$.
	In the context of the wireless cable method, the performance of the system is primarily determined by the isolation among the established wireless links, which is computed based on $\left |\mathbf{C} \cdot \hat{\mathbf{C}}^{-1} \right | $, with higher values being desirable. A statistical analysis presented in \cite{zhang2020achieving} indicates that for a $4 \times 4$ matrix $\mathbf{C}$, an isolation of merely $10$ dB is achieved at a condition number of $15$, whereas a condition number of $20$ renders the system essentially unusable.
	Furthermore, two practical measurement examples in \cite{wangAchievingWirelessCable2025} demonstrate that the condition number for a $4 \times 4$ matrix $\mathbf{C}$ can readily reach $15.5$ and $31.8$. Moreover, as the condition number increases with the expansion of the array scale, realizing a large-scale wireless cable method for massive antenna arrays constitutes the primary challenge of the proposed framework, for which a solution will be elaborated upon in the subsequent section.
	
	\section{Large-Scale Wireless Cable Establishment: Principle and Experimental Validaiton}
	In this section, we investigate the methodology for establishing large-scale wireless cable links for massive antenna arrays and design a series of experiments to validate the proposed principle.
	\subsection{Principle}
	To achieve a low condition number for $\mathbf{C}$ within the proposed framework, the primary DoF lies in designing the configuration and physical placement of the OTA probe array.
	We first outline the design principles for a large-scale wireless cable system as illustrated in \textbf{{Principle 1}} and depicted in the upper portion of Fig. \ref{FIG_Framework} (c). 
	
	\textbf {{Principle 1}} ({Sufficient Conditions for Large-Scale Wireless Cable System}) \textit{To achieve a low condition number for the propagation matrix $\mathbf{C}$, the following three conditions are sufficient for the large-scale wireless cable system.}
	\begin{enumerate}
		\item \textit{\textbf{Probe Configuration:} The OTA probe array is preferably configured identically to the DUT antenna array. }
		\item \textit{\textbf{Placement:}  The OTA probe array must be positioned directly facing and in close proximity to the DUT antenna array.}
		\item \textit{\textbf{Probe Pattern:} The radiation patterns of the OTA probe elements are preferably identical and as narrow as possible.}
	\end{enumerate}
	
	Subsequently, we investigate the conditions requisite for ensuring that the propagation matrix $\mathbf{C}$ is well-conditioned and elucidate why \textbf{{Principle 1}} facilitates the satisfaction of these conditions.
	The core idea is for $\mathbf{C}$ to become a SDD matrix with significant dominance, while ensuring that the moduli of its diagonal elements are approximately equal.
	We define $\mathbf{C} = [c_{ij}]$, where $c_{ij}$ denotes the transmission coefficient from the $j$-th Tx element to the $i$-th Rx element. 
	An SDD matrix $\mathbf{C}$ is defined such that for all $i$ and $j$, the following condition is satisfied:
	\begin{equation} 
		\left|c_{i i}\right| > \sum_{j \neq i}\left|c_{i j}\right|.
	\end{equation}
	According to the Gerschgorin circle theorem \cite{varga2011gervsgorin}, SDD ensures that the matrix is non-singular.
	We derive the upper bound of the condition number of $\mathbf{C}$ based on the definition of the infinity norm, given by
	\begin{equation} \label{eq:cond-infinity-C}
		\kappa_{\infty}(\mathbf{C})=\|\mathbf{C}\|_{\infty}\left\|\mathbf{C}^{-1}\right\|_{\infty}.
	\end{equation}
	First, we can readily obtain $\|\mathbf{C}\|_{\infty}$, given by
	\begin{equation} \label{eq:C-infinity-norm}
		\|\mathbf{C}\|_{\infty} = \max _{i} \sum_{j=1}^{n}\left|c_{i j}\right| = \max _{i}\left(\left|c_{i i}\right|+\sum_{j \neq i}\left|c_{i j}\right|\right).
	\end{equation}
	Then, utilizing the Ahlberg--Nilson--Varah bound for SDD matrices \cite{varah1975lower}, the infinity norm of the inverse matrix $\mathbf{C}^{-1}$ satisfies the following:
	\begin{equation}  \label{eq:C-inv-infinity-norm-bound}
		\left\|\mathbf{C}^{-1}\right\|_{\infty} \leq \frac{1}{\min  _{i}\left(\left|c_{i i}\right|-\sum_{j \neq i}\left|c_{i j}\right|\right)}.
	\end{equation}
	Combining (\ref{eq:cond-infinity-C}), (\ref{eq:C-infinity-norm}), and (\ref{eq:C-inv-infinity-norm-bound}), we can obtain the upper bound of the condition number of $\mathbf{C}$, given by
	\begin{equation}  \label{eq:cond-infinity-C-upper-bound}  
		\kappa_{\infty}(\mathbf{C})= \|\mathbf{C}\|_{\infty}\left\|\mathbf{C}^{-1}\right\|_{\infty}\leq	 \frac{\max  _{i}\left(\left|c_{i i}\right|+\sum_{j \neq i}\left|c_{i j}\right|\right)}{\min  _{i}\left(\left|c_{i i}\right|-\sum_{j \neq i}\left|c_{i j}\right|\right)}.
	\end{equation}
	Similar steps to the aforementioned derivation process can also be found in \cite{Chen2022Infinity}.
	Next, we define $d_{\mathrm{max}} = \max_{i} \left| c_{ii} \right|$ and $d_{\mathrm{min}} = \min_{i} \left| c_{ii} \right|$ as the maximum and minimum moduli of the diagonal elements of matrix $\mathbf{C}$, respectively.
	And we define that the ratio of the sum of the moduli of the off-diagonal elements relative to the diagonal element is controlled by $\epsilon$, i.e., $\sum_{j \neq i}\left|c_{i j}\right| \leq \epsilon\left|c_{i i}\right|$, where $0 \leq \epsilon < 1$. A smaller $\epsilon$ indicates stronger diagonal dominance.
	Then, we can derive the following inequalities:
	\begin{equation}
		\begin{array}{l}\label{eq:C-norm-epsilon-bound}
			\max _{i}\left(\left|c_{i i}\right|+\sum_{j \neq i}\left|c_{i j}\right|\right) \leq \max _{i}\left|c_{i i}\right|(1+\epsilon)=d_{\max }(1+\epsilon), \\
			\min _{i}\left(\left|c_{i i}\right|-\sum_{j \neq i}\left|c_{i j}\right|\right) \geq \min _{i}\left|c_{i i}\right|(1-\epsilon)=d_{\min }(1-\epsilon).
		\end{array}
	\end{equation}
	By combining (\ref{eq:cond-infinity-C-upper-bound}) and (\ref{eq:C-norm-epsilon-bound}), we can derive the upper bound for the condition number of SDD matrix $\mathbf{C}$, given by
	\begin{equation} \label{eq:cond-infinity-C-epsilon-upper-bound}
		\kappa_{\infty}(\mathbf{C}) \leq \frac{d_{\max }(1+\epsilon)}{d_{\min }(1-\epsilon)}
		=
		\left(\frac{d_{\max }}{d_{\min }}\right) \cdot \frac{1+\epsilon}{1-\epsilon}.
	\end{equation}
	From (\ref{eq:cond-infinity-C-epsilon-upper-bound}), it can be deduced that minimizing the upper bound of $\kappa_{\infty}(\mathbf{C})$ involves two factors. First, $\frac{d_{\max }}{d_{\min }}$ should be minimized. This requires that the variation in the moduli of the diagonal elements of matrix $\mathbf{C}$ be small, such that $\frac{d_{\max }}{d_{\min }}$ approaches 1. 
	Second, $\frac{1+\epsilon}{1-\epsilon}$ should be minimized. Since this expression is monotonically increasing with $\epsilon$, this requires $\epsilon$ to be as small as possible. According to the definition of $\epsilon$, a smaller $\epsilon$ indicates that matrix $\mathbf{C}$ exhibits stronger diagonal dominance. 
	Finally, the above conclusion is summarized in \textbf{Criterion 1}.
	
	\textbf{Criterion 1} ({Sufficient Conditions for a Low Condition Number Matrix}) \textit{To achieve a low condition number for the matrix $\mathbf{C}$, the following three conditions are sufficient.}
	\begin{enumerate}
		\item \textit{$\mathbf{C}$ is a SDD matrix.}
		\item \textit{$\mathbf{C}$ exhibits significant diagonal dominance.}
		\item \textit{The difference between the maximum and minimum moduli of the principal diagonal elements of $\mathbf{C}$ is minimal.}
	\end{enumerate}
	
	We now elucidate why the design principle for the OTA probe array, as outlined in \textbf{Principle 1}, satisfy the three sufficient conditions stipulated in \textbf{Criterion 1}, thereby ensuring that the propagation matrix $\mathbf{C}$ exhibits a low condition number.
	According to the definition of the propagation matrix $\mathbf{C}$, the elements on its principal diagonal $c_{ii}$ represent the transmission coefficients between elements of the DUT array and the OTA probe array that share the same index.
	Therefore, if the OTA probe array and the DUT array share an identical configuration and are positioned face-to-face such that elements with the same index are aligned along the direction of their respective main lobes, it can be ensured that $\mathbf{C}$ is a SDD matrix.
	Under the aforementioned conditions, positioning the OTA probe array as close as possible to the DUT array, while ensuring that the probe radiation patterns are as narrow as possible, guarantees that $\mathbf{C}$ exhibits strong diagonal dominance. Furthermore, ensuring that the OTA probes have identical radiation patterns serves to minimize discrepancies among the principal diagonal elements.
	Therefore, adhering to \textbf{Principle 1} in the wireless cable method test setup is expected to yield a propagation matrix $\mathbf{C}$ with a low condition number \footnote{It is noted that \textbf{Principle 1} constitutes a sufficient condition for achieving a propagation matrix with a low condition number. Although alternative schemes exist, the discussion thereof falls outside the scope of this paper.}.
	
	The design of the transfer matrix in wireless cable systems has been largely overlooked in the literature. We exploit this DoF to render the wireless cable method suitable for large-scale DUTs. Failure to meet these conditions may results in a deterioration of the condition number, which will be experimentally validated in the next subsection.
	This solution holds significant promise for practical applications due to its compactness resulting from the small measurement distance, the feasibility of employing identical probe and DUT array designs, and the elimination of the requirement for an anechoic chamber. In contrast, alternative solutions typically necessitate anechoic environments, large measurement distances, or specialized facilities such as compact antenna test ranges (CATR) \cite{tancioni2019over} or plane wave generators (PWG) \cite{yu2025plane}. Consequently, our proposed solution offers a cost-effective alternative.
	
	\subsection{Experimental Validaiton}
\begin{figure}[!t]
		\centering
		\subfloat[]{\includegraphics[width=0.45\textwidth]{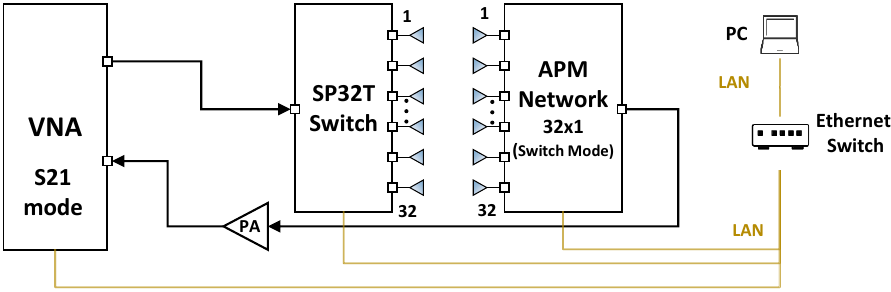}}
		\hfill 	
		\subfloat[]{\includegraphics[width=0.38 \textwidth]{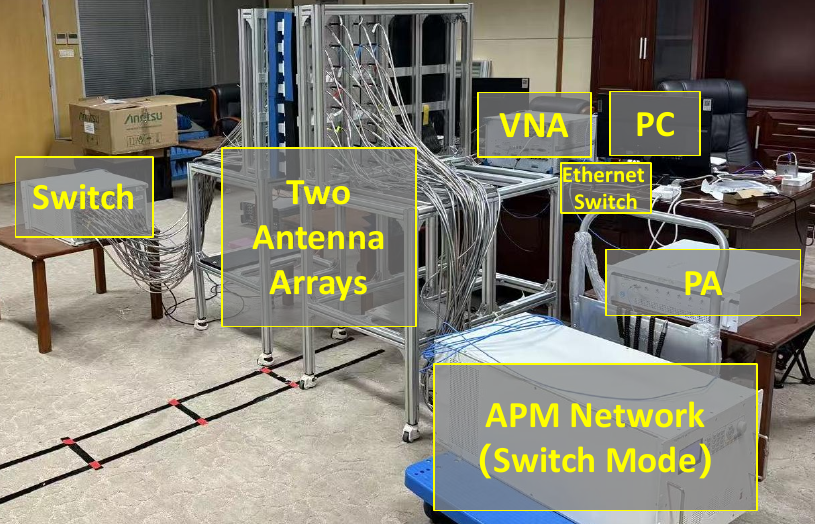}}
		\caption{Experimental setup for investigating the characteristics of the propagation matrix. (a) Illustration of the measurement setup. (b) Photograph of the setup under laboratory conditions.}
		\label{setup_wc_1}
	\end{figure}

	To validate the proposed principle, we designed the experimental setup illustrated in Fig. \ref{setup_wc_1}, which consists of the following components:
	\begin{enumerate}
		\item The VNA was configured in $S_{21}$ mode to acquire the channel response at a frequency of 3.5 GHz.
		\item Two $4\times8$ UPAs were adopted to transmit and receive signals, mimicking the ISAC BS antenna array and the probe array. These two UPAs share an identical design. In this section, we employ two pairs of identical UPA designs, specifically a waveguide antenna array and a patch antenna array.
		\item The Topyoung TSS-C601 switch box was configured as a  single-pole-32-throw (SP32T) switch.
		\item The Topyoung MCS2350A2-64B16 APM network was configured in switch mode as an SP32T switch.
		\item A power amplifier (PA) was utilized to amplify the signal to compensate for the insertion loss of the APM network.
	\end{enumerate}
	
	Based on the setup shown in Fig. \ref{setup_wc_1}, we designed three distinct experiments to verify whether the propagation matrix adheres to the proposed principles at different distances and with different element radiation patterns, as well as to investigate the condition number for  $128 \times 128$ large-scale $\mathbf{C}$.
	
	\subsubsection{Condition Number of \texorpdfstring{$\mathbf{C}$}{C} at Different Inter-Array Distances} 
\begin{figure}[!t]
		\centering
		\includegraphics[width=0.42\textwidth]{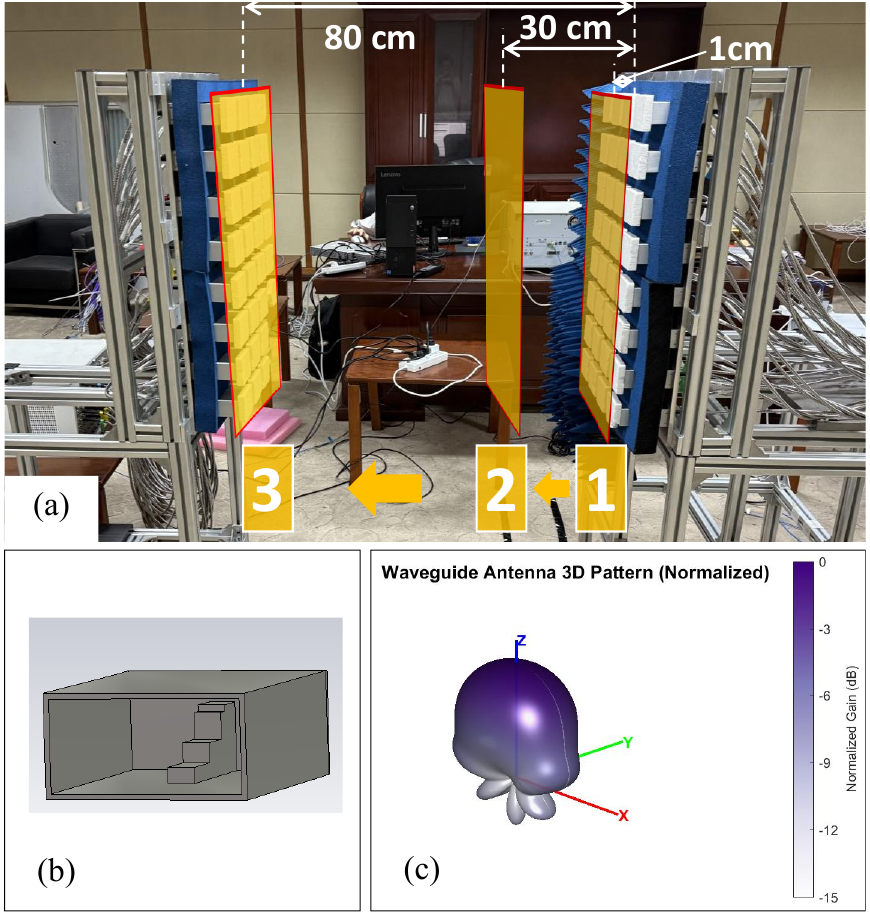} 
		\caption{Experiment setup for different inter-array distances. (a) Photograph of the experimental setup. (b) Schematic of the waveguide antennas. (c) Radiation pattern of the waveguide antennas.}
		\label{fig_distance} 
	\end{figure}

\begin{figure}[!t]
		\centering
		\begin{tabular}{cc}
			\includegraphics[width=0.46\linewidth]{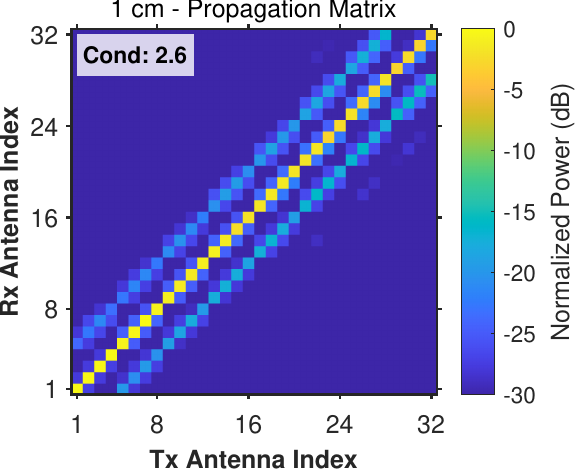} &
			\includegraphics[width=0.46\linewidth]{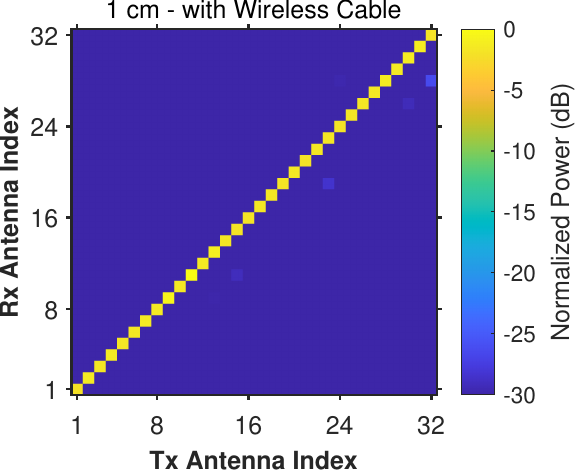} \\
			\footnotesize (a) & \footnotesize (b) \\[0.1cm]
			
			\includegraphics[width=0.46\linewidth]{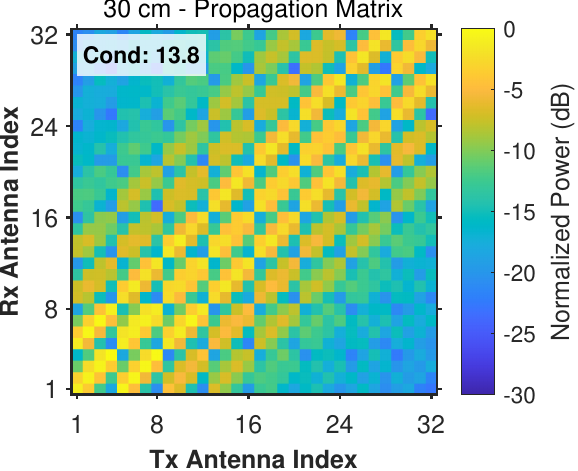} &
			\includegraphics[width=0.46\linewidth]{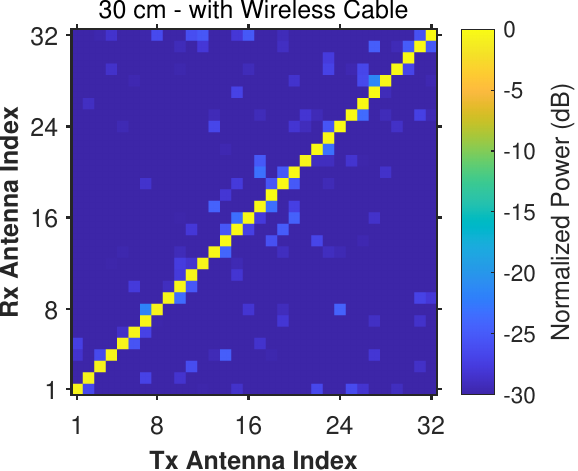} \\
			\footnotesize (c)  & \footnotesize (d)  \\[0.1cm]
			
			\includegraphics[width=0.46\linewidth]{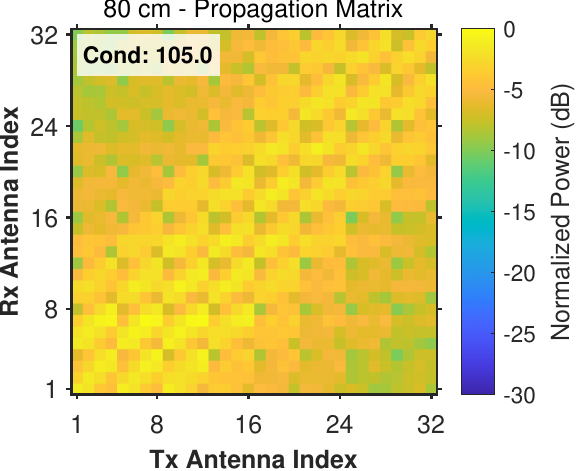} &
			\includegraphics[width=0.46\linewidth]{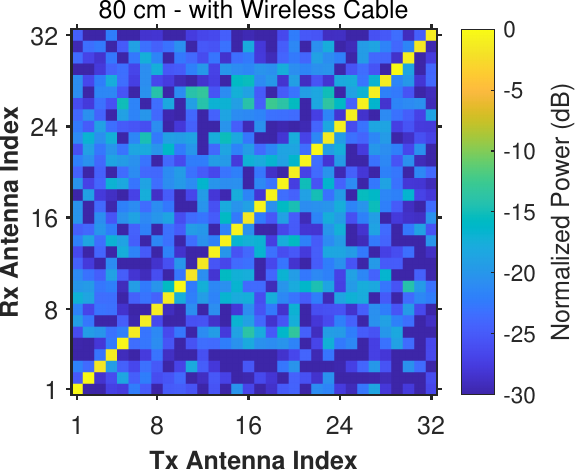} \\
			\footnotesize (e)  & \footnotesize (f) 
		\end{tabular}
		\caption{Visualization of the measured channel matrices. (a)-(f) show the comparison across different distances.}
		\label{fig:distance_comparison}
	\end{figure}

	We positioned two $4\times8$ UPAs composed of waveguide antennas with narrow beamwidths face-to-face, ensuring that probes with identical indices on both arrays were directly aligned. The schematic and radiation pattern of the waveguide antennas are illustrated in Fig. \ref{fig_distance} (b) and Fig. \ref{fig_distance} (c), respectively. Based on this setup, we varied the distance between the arrays to 1~cm, 30~cm, and 80~cm, as shown in Fig. \ref{fig_distance} (a).
	The measured $\hat{\mathbf{C}} \in \mathbb{C}^{32 \times 32} $ varies with inter-array distances as shown in Figs. \ref{fig:distance_comparison} (a), (c), and (e). It can be observed that each $\hat{\mathbf{C}}$ exhibits certain diagonal dominance characteristics. As the distance decreases from 80 cm to 30 cm and further to 1 cm, the diagonal dominance becomes increasingly significant, with the condition number correspondingly reducing from $105$ to $13.8$ and finally to $2.6$, respectively. This observation is highly consistent with the theoretical analysis presented earlier.

	Subsequently, we inverted the acquired $\hat{\mathbf{C}}$ to obtain the calibration matrix $\hat{\mathbf{C}}^{-1}\in \mathbb{C}^{32 \times 32} $, which was then imported into the APM network to establish wireless cable link. The propagation matrix between the arrays was remeasured to finally obtain $\mathbf{C} \cdot \hat{\mathbf{C}}^{-1}\in \mathbb{C}^{32 \times 32} $ for each distance.
	As shown in Figs. \ref{fig:distance_comparison} (b), (d), and (f), the isolation between the wireless cables progressively improves as the inter-array distance decreases, where a deeper blue color off the diagonal indicates higher isolation of the wireless cable link. At a distance of 1 cm, the average isolation exceeds $30$ dB, indicating that the quality of the established wireless links satisfies the testing requirements. Conversely, at a distance of 80 cm, the isolation degrades to as low as $5$ dB. This is because a higher condition number amplifies the non-ideal factors of the test system, particularly those arising from imperfections in the APM network, such as quantization errors, stepping errors, and inter-channel inconsistencies.

	\begin{figure}[!t]
		\centering
		\includegraphics[width=0.40\textwidth]{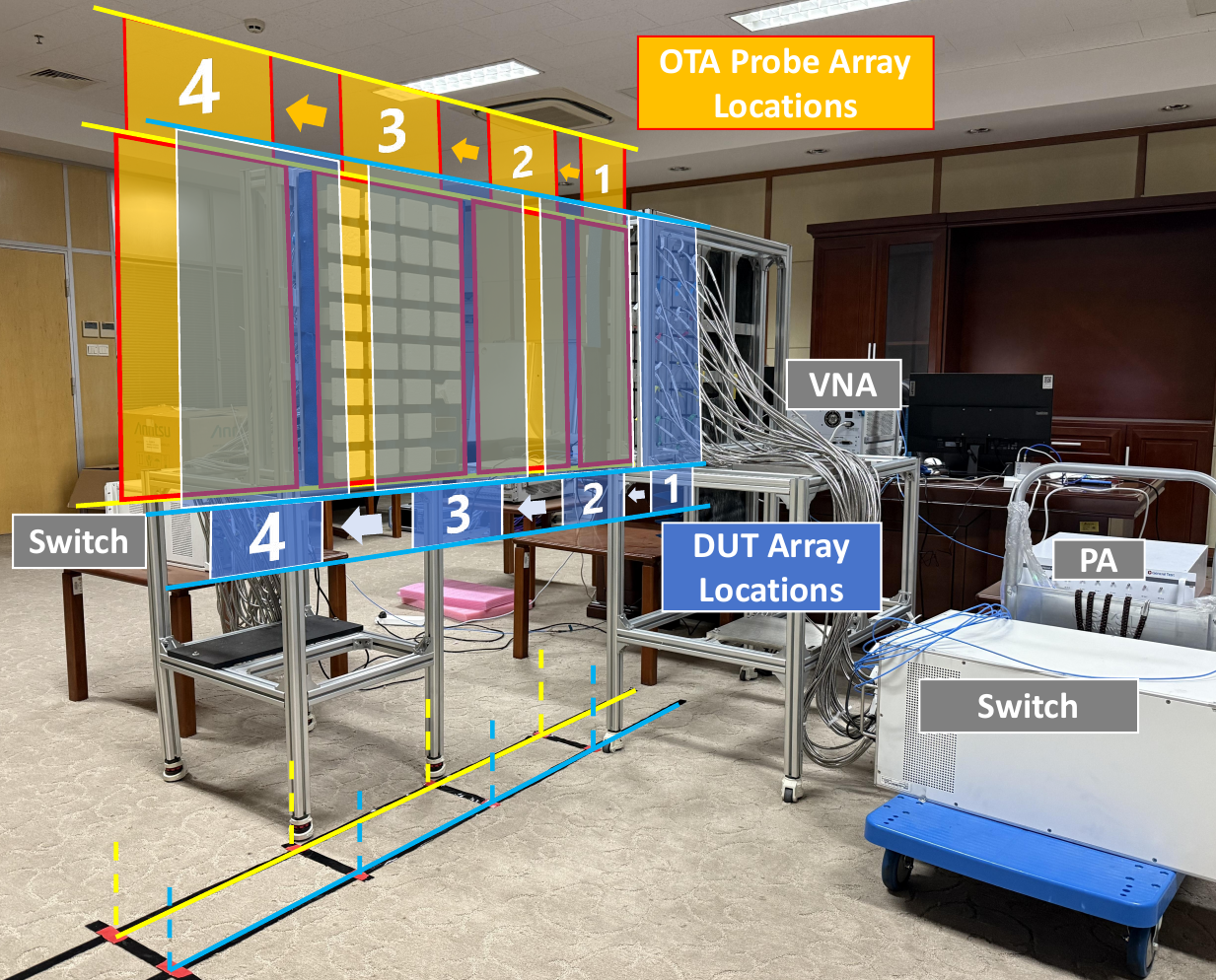} 
		\caption{Photograph of the virtual synthetic array experiment.
		}
		\label{fig_syn_photo} 
	\end{figure}

	\begin{figure}[!t]
		\centering
		\includegraphics[width=0.355\textwidth]{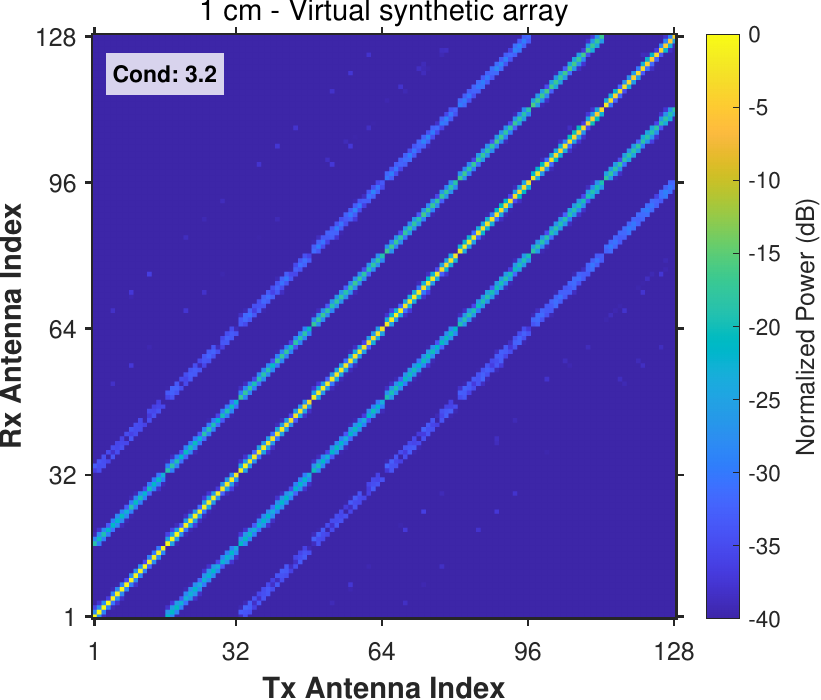} 
		\caption{Visualization of the synthesized matrix $\hat{\mathbf{C}}^{syn}_{128 \times 128}$. 
		}
		\label{fig_syn_result} 
	\end{figure}

	\subsubsection{Condition Number of Large-Scale \texorpdfstring{$\mathbf{C}$}{C}}
	To validate whether the proposed principle remains effective for large-scale arrays, we employed a virtual synthetic array approach to acquire a synthetic propagation matrix $\hat{\mathbf{C}}^{syn}_{{128 \times 128}}\in \mathbb{C}^{128 \times 128}$, as shown in Fig. \ref{fig_syn_photo}.
	Two UPA arrays are positioned with an inter-array distance of $1$ cm, and each possesses four virtual sampling locations, denoted by orange and blue rectangles in the figure, respectively, accompanied by numbering labels.The positions indicated by the orange rectangles are designated as the locations for the OTA probe array, while the blue rectangles represent the locations for the DUT array. The OTA probe array traverses through the numbered positions $1\sim 4$. For each position occupied by the OTA probe array, the DUT array similarly traverses its corresponding numbered positions $1\sim 4$. At each configuration, the propagation matrix between the arrays is recorded. Consequently, a total of $4 \times 4 = 16$ sub-matrices $\hat{\mathbf{C}}^{sub} \in \mathbb{C}^{32 \times 32}$ are measured. Finally, $\hat{\mathbf{C}}^{syn}_{{128 \times 128}}$ is synthesized via a virtual synthetic post-processing algorithm.
	As shown in Fig. \ref{fig_syn_photo}, the synthesized matrix $\hat{\mathbf{C}}^{syn}_{128 \times 128}$ continues to exhibit the characteristics of a SDD matrix with strong diagonal dominance, and the moduli of its diagonal elements remain similar. 
	Consequently, even with the matrix dimensions reaching $128 \times 128$, the condition number remains as low as $3.2$, which satisfies the requirements for establishing high-quality wireless cable links, thereby validating the effectiveness of \textbf{Principle 1}.
	This experiment demonstrates that the proposed scheme is scalable to MIMO systems, rendering it highly valuable for OTA testing of future BSs, particularly in the context of the emerging extremely large antenna arrays or gigantic MIMO systems. To the best of our knowledge, this represents the only viable solution for the OTA testing of massive MIMO BSs.

	\begin{figure}[!t]
		\centering
		\includegraphics[width=0.49\textwidth]{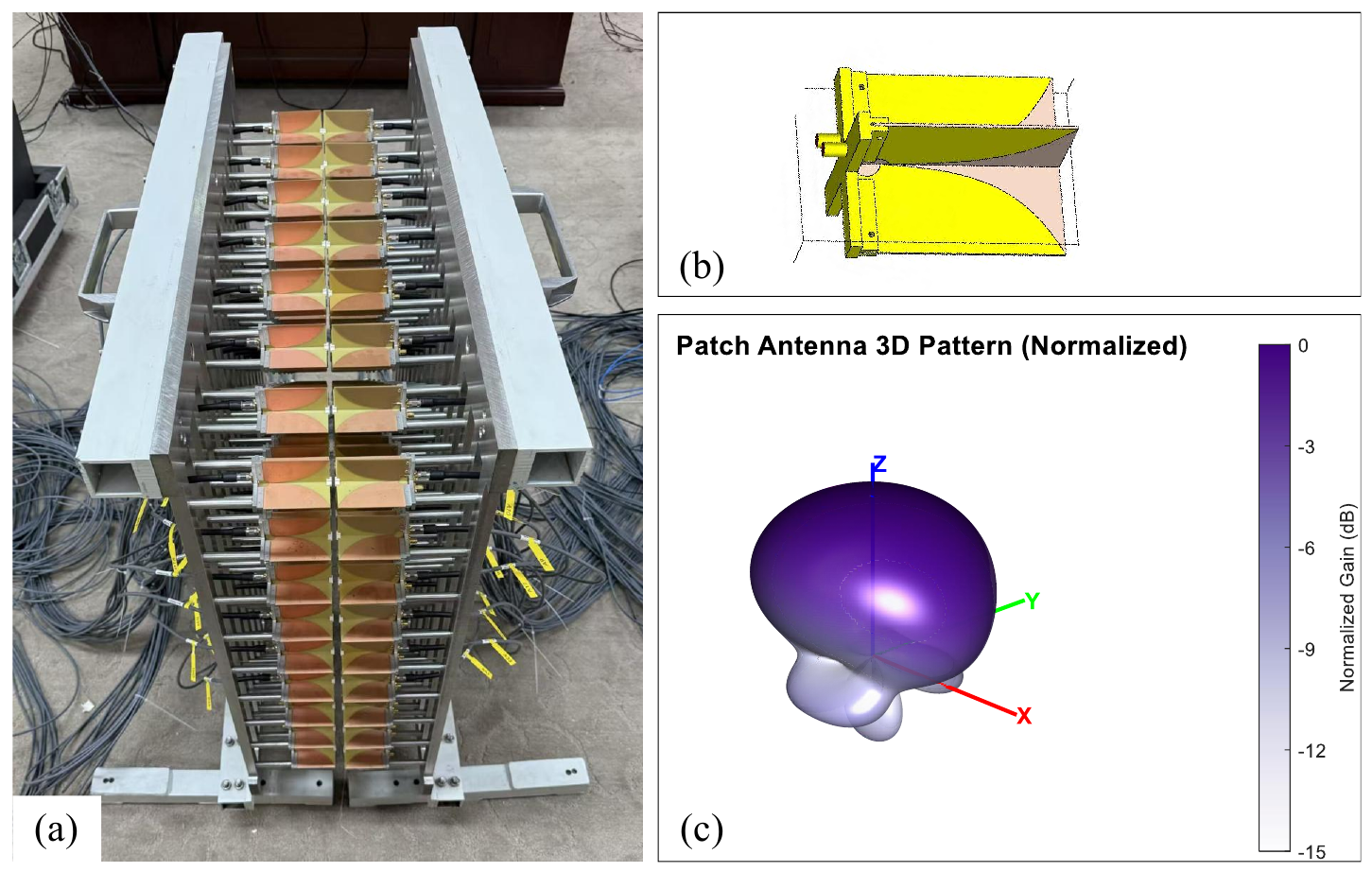} 
		\caption{Experiment setup for different inter-array distances. (a) Photograph of the experimental setup. (b) Schematic of the waveguide antennas. (c) Radiation pattern of the waveguide antennas.
		}
		\label{fig_wide_probe} 
	\end{figure}

	\begin{figure}[!t]
		\centering
		\includegraphics[width=0.3\textwidth]{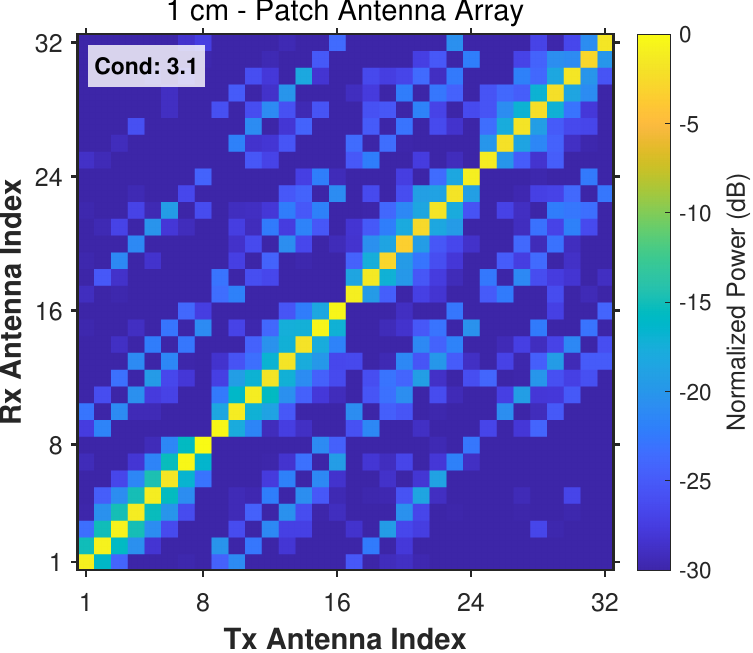} 
		\caption{Schematic of the propagation matrix between patch arrays with an inter-array distance of 1 cm.}
		\label{fig_wide_result} 
	\end{figure}
	
	\subsubsection{Condition Number of \texorpdfstring{$\mathbf{C}$}{C} with Wider Pattern Probes}
	According to the previous discussion, employing probes with wider patterns is expected to diminish the diagonal dominance of $\mathbf{C}$. To verify this, patch antennas featuring wider patterns were adopted; the schematic of the antenna structure and its radiation pattern are illustrated in Fig. \ref{fig_wide_probe} (b) and Fig. \ref{fig_wide_probe} (c), respectively.
	As shown in Fig. \ref{fig_wide_probe} (a), we positioned two UPAs equipped with patch antennas at a distance of 1 cm to acquire $\hat{\mathbf{C}}$. The measured results are presented in Fig. \ref{fig_wide_probe} (a). It can be observed that, compared to the test based on waveguide antennas shown in Fig. \ref{fig:distance_comparison} (a), the $\hat{\mathbf{C}}$ obtained using patch antennas exhibits  weaker diagonal dominance, and the condition number reaches $3.1$. Nevertheless, it remains within the usable range, thereby validating the previously discussed criterion.

	\section{Multi-Target Emulation Experiment}
	In this section, we design an ISAC BS sensing scenario involving multiple drone targets and perform a comparative validation between the traditional conducted setup and the OTA setup proposed in this work.
	\begin{figure}[!t]
		\centering
		\includegraphics[width=0.4\textwidth]{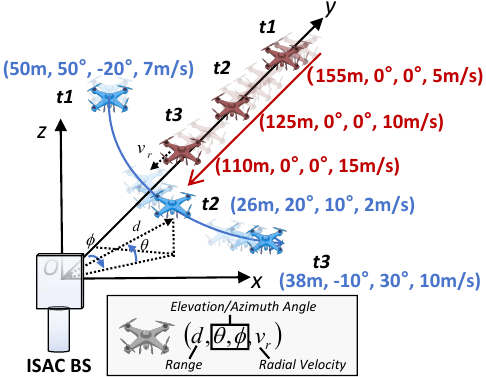} 
		\caption{Target sensing scenario and coordinate system configuration.  We define the ISAC BS array center as the origin, with the vertical array plane facing the positive $y$-axis, and the $z$-axis pointing upward. The positive direction of the azimuth angle $\phi$ increases along the positive $x$-axis from the positive $y$-axis. The elevation angle $\theta$ increases in the direction of the positive $z$-axis from the $xoy$-plane.
		}
		\label{fig_tar} 
	\end{figure}
	\subsection{Target Sensing Scenario}

	A sensing scenario is designed wherein an ISAC BS detects two drone targets, as illustrated in Fig. \ref{fig_tar}. Each drone is modeled as a far-field point target. Across three channel snapshots $t_1$, $t_2$, and $t_3$, the two drones exhibit distinct ranges, elevation angles $\theta$, azimuth angles $\phi$, and radial velocities relative to the BS. Drone 1 (blue) traverses a curved trajectory, while Drone 2 (red) approaches the ISAC BS along the negative $y$-axis from a distant initial position. The parameters characterizing this target sensing scenario are summarized in Table \ref{tab:drone_measurements}. 
	Although the parameters employed in this target sensing scenario are not derived from sophisticated electromagnetic simulation tools and thus may not precisely replicate real-world conditions, this discrepancy does not compromise the validity of verifying the capability of the proposed framework for accurate sensing target emulation.

	\begin{table}
		\caption{Measurement Equipment and Parameters}
		\label{Mea_parameter1}
		\centering
		\renewcommand\arraystretch{1.1}
		\begin{tabular}{@{}ll@{}}
			\toprule
			\textbf{Equipment/Parameter}          & \textbf{Value}         
			\\ 
			\midrule
			\underline{CE}             & Keysight Propsim F32    \\
			CE bandwidth                 & 40 MHz                  \\
			CE center frequency                    & 3500 MHz                  \\ 
			\underline{VNA}                          & Ceyear 3671C     \\
			VNA center frequency                  & 3500 MHz                \\
			VNA bandwidth                         & 40 MHz                  \\
			VNA IF bandwidth                         & 2000 Hz                  \\
			Number of frequency points  $N_f$           & 1001                   \\
			VNA Power                         & 20 dBm               \\
			\underline{Power Amplifier}              & General Test PA M80706A 
			\\ 
			\underline{Switch}              & Topyoung TSS-C601 
			\\ 
			\underline{APM network}              & Topyoung MCS2350A2-64B16
			\\
			APM	phase-shifter bit resolution          & 10 bit
			\\
			\bottomrule
		\end{tabular}
	\end{table}
	\begin{figure}[!t]
		\centering
		\subfloat[]{\includegraphics[width=0.48\textwidth]{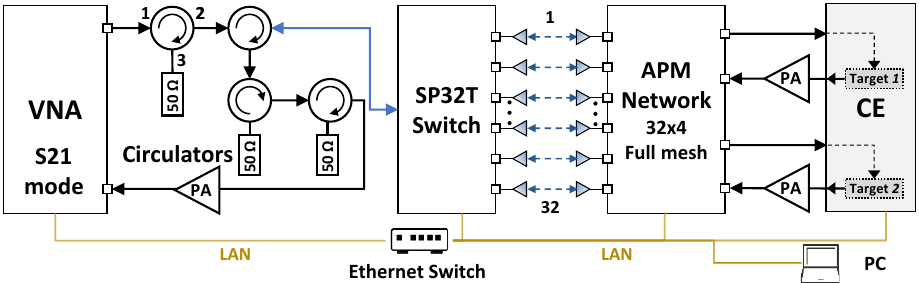}}
		\hfill 	
		\subfloat[]{\includegraphics[width=0.42 \textwidth]{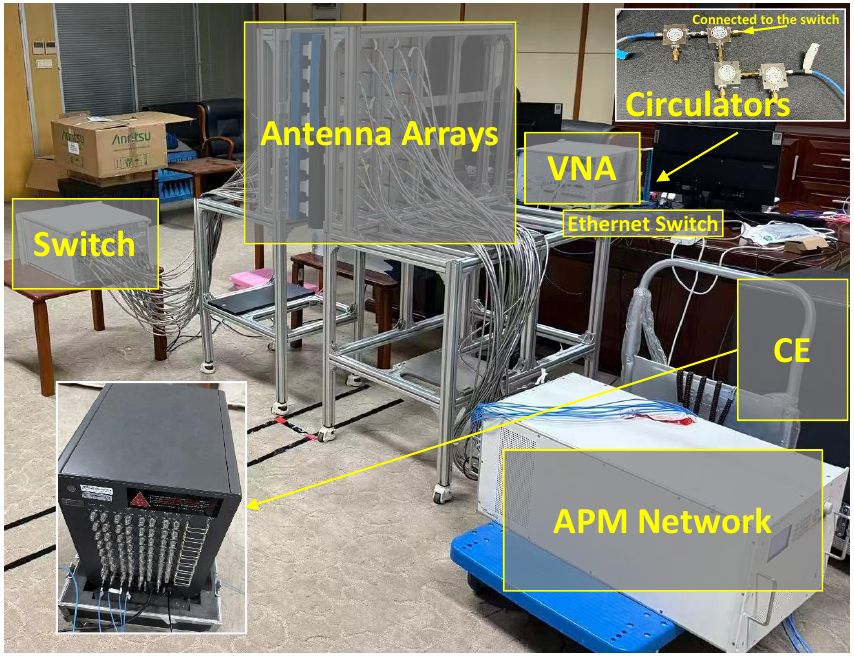}}
		\caption{Experiment setup of the sensing targets OTA  emulation for the ISAC BS. (a) Illustration of the measurement setup. (b) Photo of the setup in the laboratory condition.}
		\label{setup_pM}
	\end{figure}
	
	\subsection{Experiment Setup}
	As shown in Fig. \ref{setup_pM}, the OTA experimental validation setup comprises both the ISAC BS and the testing system:
	
	$\bullet$ 
	To emulate an ISAC BS operating in ADTR sensing mode, a vector network analyzer (VNA), a single-pole 32-throw (SP32T) switch, four cascaded circulators ensuring high transceiver isolation, and a $4 \times 8$ UPA waveguide antenna array are employed. This configuration emulates an ISAC BS  operating at 3.5 GHz with half-wavelength element spacing. The high transceiver isolation provided by the cascaded circulator configuration enables simultaneous transmission and reception signal recording, thereby accurately mimicking the operation of the ISAC BS in ADTR mode. The VNA and switching system are utilized to acquire the sensing channel matrix, with subsequent array signal processing performed via offline post-processing algorithms.
	
	$\bullet$ 
	For the emulation system, the APM network is configured with a $32 \times 4$ full-mesh architecture wherein 32 Type-A ports are connected to a $4 \times 8$ UPA waveguide antenna array identical to that of the emulated ISAC BS and four Type-B ports are connected to the CE. Serving as the radar target simulator, the Keysight PROPSIM F32 CE is configured to emulate two targets utilizing four RF ports while consuming only two internal channel resources. The receiving array steering vector $\mathbf{a}_{r}(\mathbf{\Theta}_{n}(t))$  is loaded into the APM network while the channel gain $G_{n}(t, \mathbf{\Theta}_{n}(t))$, Doppler shift $\nu_{n}(t)$, and delay $\tau_{n}(t)$ are loaded into the CE. The channel impulse response (CIR) files loaded into the CE are configured for Drone 1 and Drone 2 with their corresponding delays, power levels, and radial velocities.
	The update rate of the CIR files is configured to meet the required fading resolution with a time interval of 1.43 ms between consecutive samples, and each snapshot consists of $N_t = 1000$ CIR samples. For velocity estimation, the CE pauses at each CIR sample; after the VNA measures the corresponding CFR over $N_f = 1001$ frequency points within a 40 MHz bandwidth, the CE advances to the subsequent CIR sample and repeats the process until the sequence is completed. This operational methodology is widely adopted for MIMO OTA channel validation measurements \cite{3gpp-tr-38-827}. Consequently, a CFR dataset of dimensions $N_t \times N_f \times 32$ is recorded. Finally, the detailed instrument parameters for this experiment are summarized in Table \ref{Mea_parameter1}.
	
	For comparison, we conducted experiments based on a conducted target simulation system, the architecture and parameters of which can be found in \cite{liMultitargetFlexibleAngular2025}. It should be noted that the experimental results in this work may exhibit slight deviations from those in \cite{liMultitargetFlexibleAngular2025} as the data were acquired from a newly conducted experiment rather than utilizing the same dataset.
	In the OTA experimental setup, the two UPA arrays are positioned face-to-face with an inter-array distance of merely 1 cm, consistent with the configuration shown in Fig. \ref{fig_distance} (a).
	During the experiment, the positions of the two UPA arrays remain fixed to ensure the stability of the wireless cable links.

	\begin{table*}[htbp]
		\centering
		\caption{Target, Conducted, and OTA Parameters of Drones at Different Channel Snapshots}
		\label{tab:drone_measurements}
		\setlength{\tabcolsep}{5pt} 
		\begin{tabular}{@{}lc*{9}{c}@{}}
			\toprule
			\multirow{2}{*}{\textbf{Drone index}} & \multirow{2}{*}{\textbf{Parameter}} & \multicolumn{3}{c}{\textbf{Snapshot} $t_1$} & \multicolumn{3}{c}{\textbf{Snapshot} $t_2$} & \multicolumn{3}{c}{\textbf{Snapshot} $t_3$} \\
			\cmidrule(lr){3-5}\cmidrule(lr){6-8}\cmidrule(lr){9-11}
			& & \textbf{Target} & \textbf{Conducted} & \textbf{OTA} & \textbf{Target} & \textbf{Conducted} & \textbf{OTA} & \textbf{Target} & \textbf{Conducted} & \textbf{OTA} \\
			\midrule
			\multirow{6}{*}{Drone 1} & Range ($m$) & 50 & 50 & 50 & 26 & 26 & 26 & 38 & 38 & 38 \\
			& Radial Velocity ($m/s$) & 7 & 7 & 7 & 2 & 2 & 2 & 10 & 10 & 10 \\
			& Elevation Angle $\theta$ (°) & 50 & 50 & 50 & 20 & 20 & 20 & -10 & -10 & -10 \\
			& Azimuth Angle $\phi$ (°) & -20 & -20 & -20 & 10 & 10 & 10 & 30 & 30 & 30 \\
			& Normalized Channel Gain (dB) & -5 & -5 & -5 & 0 & 0 & 0 & -3 & -3 & -3 \\
			& PAS Similarity & - & 99.2\% & 97.2\% & - & 98.5\% & 92.2\% & - & 98.4\% & 93.3\% \\ 
			\midrule
			\multirow{6}{*}{Drone 2} & Range ($m$) & 155 & 155 & 155 & 125 & 125 & 125 & 110 & 110 & 110 \\
			& Radial Velocity ($m/s$) & 5 & 5 & 5 & 10 & 10 & 10 & 15 & 15 & 15 \\
			& Elevation Angle $\theta$ (°) & 0 & 0 & 0 & 0 & 0 & 0 & 0 & 0 & 0 \\
			& Azimuth Angle $\phi$ (°) & 0 & 0 & 0 & 0 & 0 & 0 & 0 & 0 & 0 \\
			& Normalized Channel Gain (dB) & -25 & -25.9 & -26.2 & -20 & -20.7 & -20.3 & -13 & -12.8 & -13.1 \\
			& PAS Similarity & - & 99.5\% & 87.4\% & - & 98.7\% & 92.8\% & - & 99.5\% & 92.4\% \\ 
			\bottomrule
		\end{tabular}
	\end{table*}

	\begin{figure}[!t]
		\centering
		\subfloat[\footnotesize Snapshot $t_1$]{
			\includegraphics[width=\linewidth]{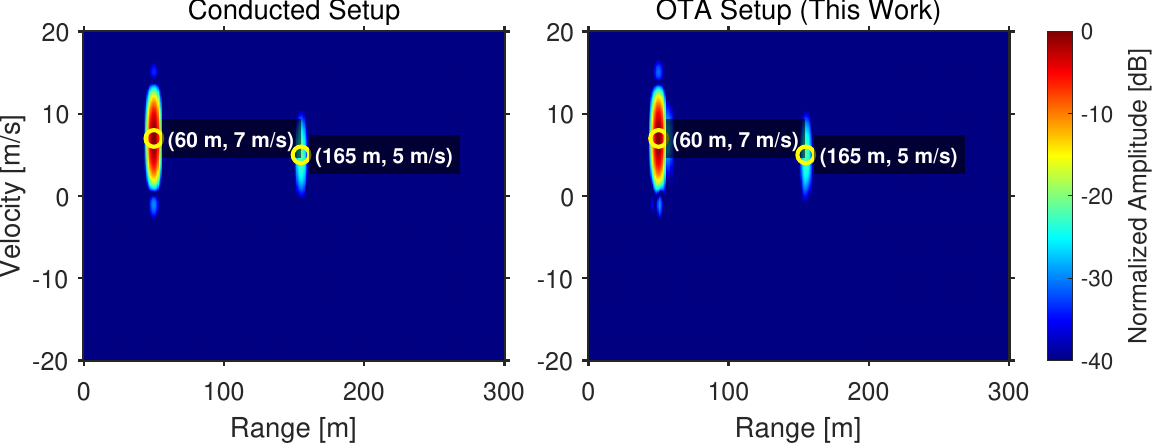}
			\label{fig:rd_t1}
		}
		
		\vspace{0.1cm} 
		
		\subfloat[\footnotesize Snapshot $t_2$]{
			\includegraphics[width=\linewidth]{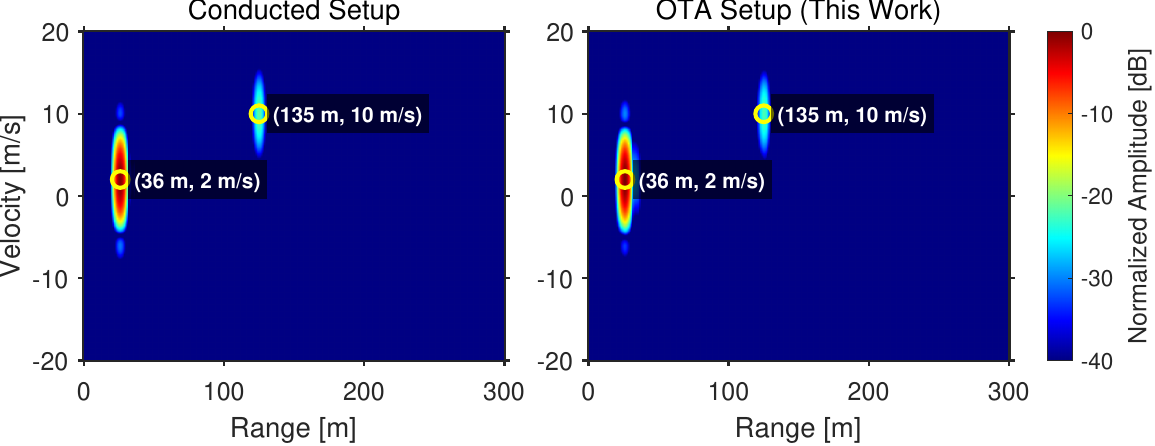}
			\label{fig:rd_t2}
		}
		
		\vspace{0.1cm} 
		
		\subfloat[\footnotesize Snapshot $t_3$]{
			\includegraphics[width=\linewidth]{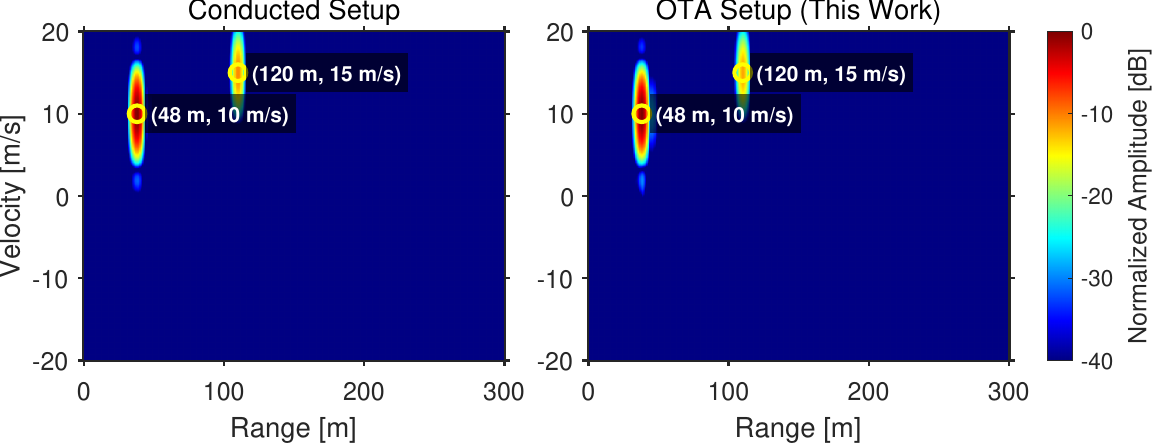}
			\label{fig:rd_t3}
		}
		
		\caption{Comparison of range-velocity maps obtained from the two setups across three distinct snapshots. (a), (b), and (c) correspond to snapshots $t_1$, $t_2$, and $t_3$, respectively.}
		\label{fig:rv_comparison}
	\end{figure}

	\subsection{Result Analysis}
	First, we perform a comparative validation of the range-velocity maps generated by the two setups for the same target scenario. We randomly select a 2D time-frequency  dataset with dimensions $N_t \times N_f$ from a single antenna port for both setups and apply a 2D Fourier transform (FT) to obtain the 2D delay-Doppler  spectrum. Finally, this spectrum is mapped to the range-velocity domain. The experimental results are presented in Fig.~\ref{fig:rv_comparison}.
	It can be observed that the proposed OTA setup yields a range-velocity map that closely approximates that of the conducted setup. Furthermore, the estimated two peaks in each map are highlighted with yellow circles, and the corresponding estimated range and velocity values are annotated. The results demonstrate that both schemes successfully estimate the parameters of multiple target drones consistent with the preset range and velocity values in each snapshot.

	\begin{figure*}[!t]
		\centering
		\begin{tabular}{cc}
			\includegraphics[width=0.48\linewidth]{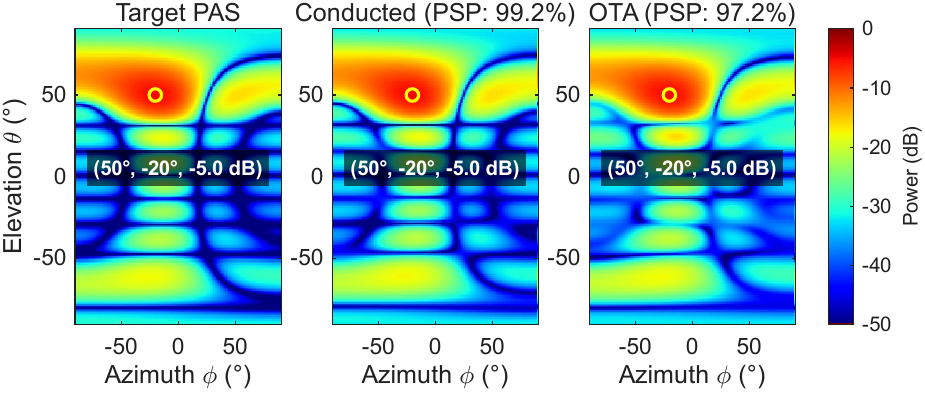} &
			\includegraphics[width=0.48\linewidth]{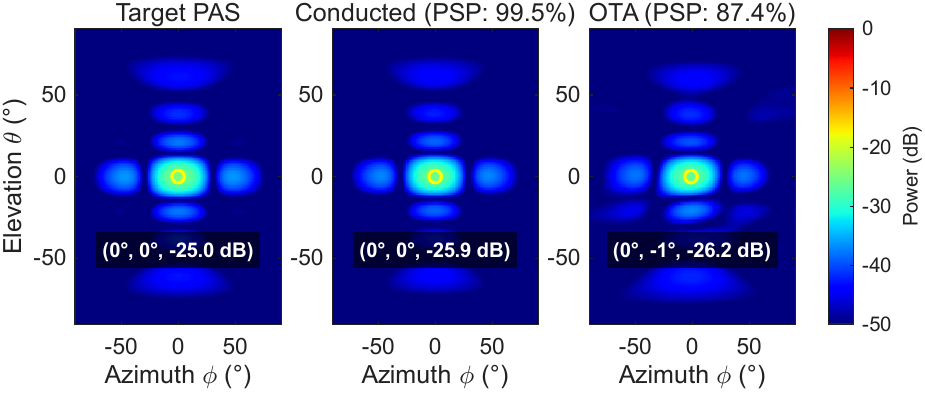} \\
			\footnotesize (a) PAS Comparison for Drone 1 at Snapshot $t_1$ & 
			\footnotesize (b) PAS Comparison for Drone 2 at Snapshot $t_1$ \\[0.2cm]
			
			\includegraphics[width=0.48\linewidth]{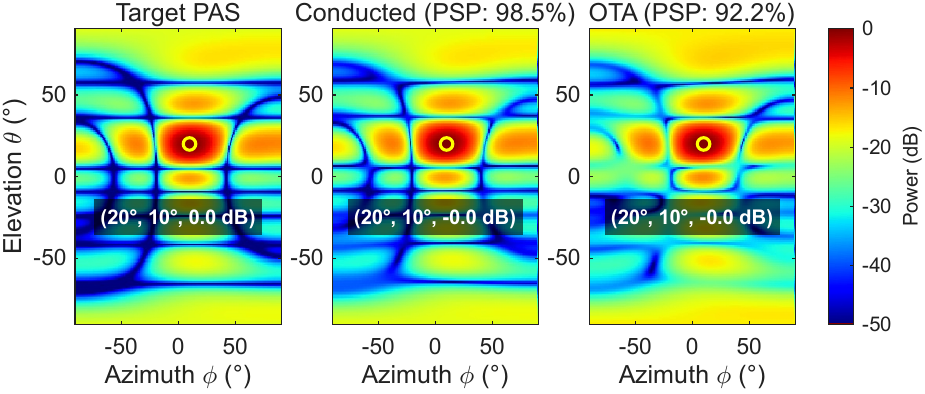} &
			\includegraphics[width=0.48\linewidth]{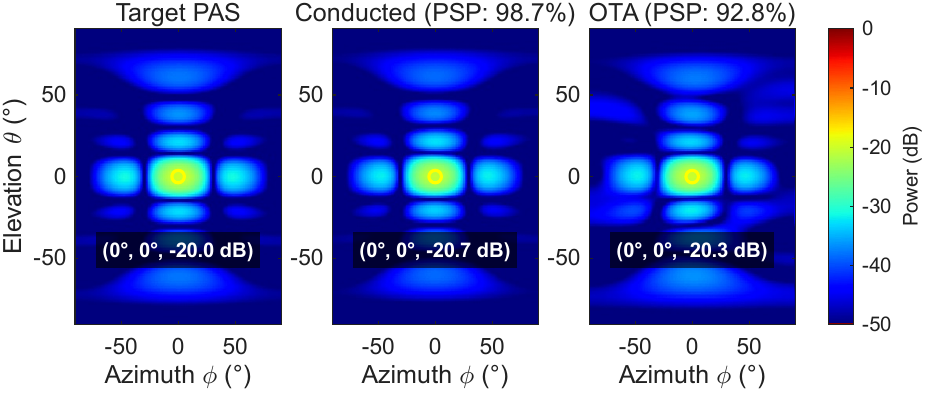} \\
			\footnotesize (c) PAS Comparison for Drone 1 at Snapshot $t_2$ & 
			\footnotesize (d) PAS Comparison for Drone 2 at Snapshot $t_2$ \\[0.2cm]
			
			\includegraphics[width=0.48\linewidth]{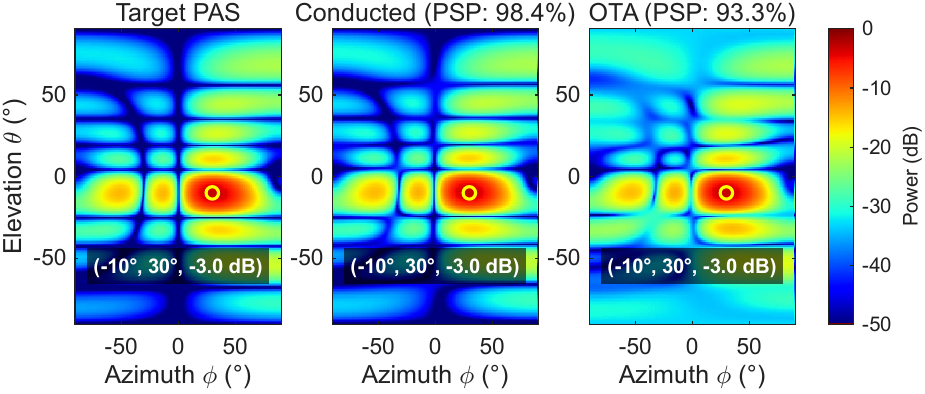} &
			\includegraphics[width=0.48\linewidth]{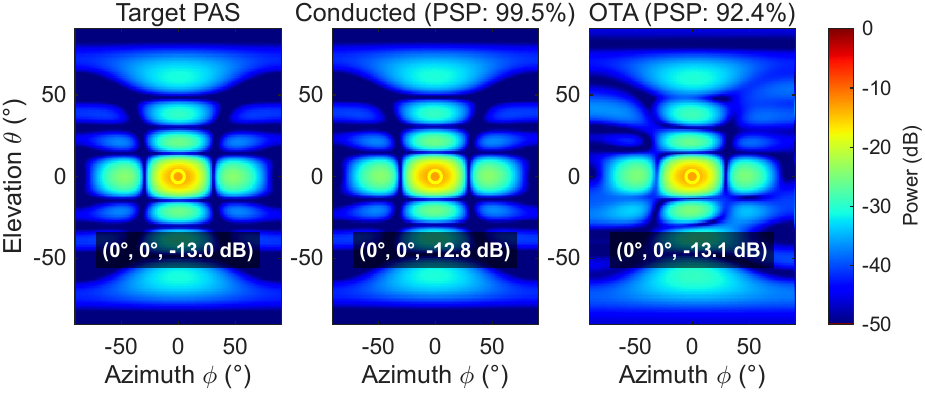} \\
			\footnotesize (e) PAS Comparison for Drone 1 at Snapshot $t_3$ & 
			\footnotesize (f) PAS Comparison for Drone 2 at Snapshot $t_3$
		\end{tabular}
		\caption{Results of the PAS across different time snapshots for different drones under various setups.}
		\label{fig:pas_results}
	\end{figure*}
	
	Subsequently, the effectiveness of spatial characteristic emulation for each target is validated through both OTA and conducted setups as well as simulations. We select the frequency-antenna dataset of dimensions $N_f \times 32$ and perform a FT to obtain the delay-antenna 2D CIR of dimensions $N_\tau \times 32$. Then, based on the estimated ranges shown in Fig. \ref{fig:rv_comparison}, we extract slices corresponding to the range of each target to acquire the CIR across the 32 array elements for the $n$-th drone at $t_n$ snapshot under each setup. Specifically, $\mathbf{h}^{OTA}_{t_n}(\tau_{n})$, $\mathbf{h}^{Con.}_{t_n}(\tau_{n})$, and $\mathbf{h}^{Target}_{t_n}(\tau_{n})$, each with dimensions $32 \times 1$, represent the  CIRs for the OTA setup, conducted setup, and simulation, respectively. 
	Finally, the power angular spectrum (PAS) for each drone and snapshot is calculated, and to quantify the PAS emulation accuracy, the PAS similarity percentage (PSP) is adopted. Readers may refer to \cite{liMultitargetFlexibleAngular2025,3gpp-tr-38-827} for the detailed calculation procedure, which is omitted here for brevity.

	As shown in Fig. \ref{fig:pas_results}, the peak of each PAS pattern is highlighted by a yellow circle, with its power value, elevation angle $\theta$, and azimuth angle $\phi$ annotated in the figure. From the estimated values, it can be observed that both the proposed OTA setup and the conducted setup accurately estimate the angles and channel gains of each drone target. For the OTA setup, the maximum angular deviation is $1$$^{\circ}$, and the maximum gain deviation is merely $1.2$ dB, both occurring for the drone at snapshot $t_1$. This is attributed to the lower preset power level in this scenario, which makes it more susceptible to noise. In terms of PAS patterns, although the reconstruction accuracy of the OTA scheme is not as high as that of the conducted setup, the PSP remains above $92\%$ in most cases, with the exception of Drone 2 at snapshot $t_1$, which registers $87.4\%$. The main lobes of the PAS diagrams for both conducted and OTA setups exhibit good reconstruction relative to the target PAS, whereas sidelobe blurring is observed, being more pronounced in the OTA scheme. In summary, Fig. \ref{fig:pas_results} demonstrates that the proposed framework can effectively emulate the spatial characteristics of multiple drone targets in an OTA manner. The experimental results of this section are summarized in Table \ref{tab:drone_measurements}.
	
	\subsection{Discussion}
	The experiments demonstrate that the proposed framework is capable of simultaneously establishing 32 stable large-scale wireless cable links and effectively emulating the range, velocity, AoA, and channel gain of multiple drone targets over the air. The successful estimation of these parameters by the emulated DUT ISAC BS validates the effectiveness of the proposed framework.
	
	It should be noted that the operational mode of the ISAC BS emulated in the experiments may not be perfectly equivalent to that of a real-world BS. For instance, this study employs a basic Bartlett beamforming algorithm, whereas actual BSs might utilize high-precision parameter estimation algorithms. Furthermore, ISAC systems may not operate exclusively in ADTR mode and could instead adopt a configuration where a subset of the array transmits while another subset receives. However, these distinctions do not compromise the validation of the principles underlying the proposed framework, particularly the verification of the principles for establishing large-scale wireless cable links.
	Moreover, to achieve high-fidelity target emulation, the calibration of inter-channel inconsistencies within the APM network and the CE as well as the intrinsic delay of the CE is essential and has been implemented in the experiments.

	\section{Conclusion}
	This paper proposes a multi-target sensing OTA emulation framework based on a large-scale wireless cable method for ISAC BS sensing test. Based on the derivation of the upper bound for the condition number of SDD matrices, design and deployment principles for OTA probe arrays targeting large-scale DUTs are proposed. Experimental validation demonstrates that the condition number of the transmission matrix for a 32-element DUT is as low as $2.6$, and remains as low as $3.2$ even for a 128-element configuration, significantly surpassing the results of previous studies. Finally, a dynamic sensing scenario involving two drones is designed to compare the proposed OTA setup with traditional conducted results. The findings indicate that the proposed scheme accurately emulates the range, velocity, AoA, and channel gain of the targets, with a maximum angular error of only $1^{\circ}$ and a maximum power error of 1.2 dB. The PSP consistently exceeds 87\%, satisfying testing requirements. 
	Compared to other potential OTA testing solutions such as CATR and PWG, the proposed system offers significant cost advantages. The experimental results from the synthetic array demonstrate its scalability for future gigantic MIMO BSs. Future work will involve validating real BSs or radars based on the proposed framework.

	\bibliographystyle{IEEEtran}
	\addcontentsline{toc}{section}{\refname}
	\bibliography{ref_WC_ISAC}
\end{document}